\documentclass[pre,showpacs,showkeys,preprintnumbers,amsmath,amssymb,superscriptaddress,twocolumn,nofootinbib]{revtex4-2}
\pdfoutput=1 
\usepackage{graphicx}
\usepackage{amsfonts}
\usepackage{url}
\usepackage{epstopdf}
\usepackage{color}
\usepackage{tikz-cd}
\usepackage{bm}
\usepackage[colorlinks=true,linkcolor=blue,citecolor=red]{hyperref}%

\def\nmax{n_{\rm max}}

\def\AA{H}
\def\AAF{\bar{F}}
\def\AAG{\bar{G}}

\def\C{\mathcal{C}}

\def\MM{{\bf M}}

\def\VV{{\bf V}}
\def\XX{{\bf X}}
\def\FF{{\bf F}}
\def\cc{{\bf c}}
\def\II{{\bf I}}

\def\pa{\partial\Omega}
\def\P{{\mathbb P}}
\def\R{{\mathbb R}}

\def\M{{\mathcal M}}

\def\X{\bm{X}}
\def\x{\bm{x}}

\begin{document}

\title{Spectral properties of the Dirichlet-to-Neumann operator for spheroids}

\author{Denis~S.~Grebenkov}
 \email{denis.grebenkov@polytechnique.edu}
\affiliation{
Laboratoire de Physique de la Mati\`{e}re Condens\'{e}e (UMR 7643), \\ 
CNRS -- Ecole Polytechnique, Institut Polytechnique de Paris, 91120 Palaiseau, France}

\date{\today}

\begin{abstract}
We study the spectral properties of the Dirichlet-to-Neumann operator
and the related Steklov problem in spheroidal domains ranging from a
needle to a disk.  An explicit matrix representation of this operator
for both interior and exterior problems is derived.  We show how the
anisotropy of spheroids affects the eigenvalues and eigenfunctions of
the operator.  As examples of physical applications, we discuss
diffusion-controlled reactions on spheroidal partially reactive
targets and the statistics of encounters between the diffusing
particle and the spheroidal boundary.
\end{abstract}





\maketitle

\section{Introduction}
\label{sec:intro}

The Dirichlet-to-Neumann operator $\M$ plays an important role in
applied mathematics, physics, and engineering.  One of its most known
applications is related to electrical impedance tomography, also known
as Calderon's problem
\cite{Cheney99,Calderon80,Borcea02,Sylvester87,Curtis91}, in which the
electric conductivity in the bulk has to be determined from electric
measurements on the boundary.  The Dirichlet-to-Neumann operator is
also employed as a ``building block'' for analyzing and solving
spectral and scattering problems in complex media via domain
decomposition (see
\cite{Agranovich,Smith96,Givoli98,Delitsyn18,Delitsyn22} and
references therein).  The eigenfunctions of $\M$ often appear as a
basis for representing and approximating harmonic functions and
related quantities
\cite{Auchmuty04,Auchmuty13,Auchmuty14,Auchmuty15,Auchmuty18}.
In chemical physics, diffusion-controlled reactions and other
diffusion-mediated surface phenomena can be described by means of the
operator $\M$ \cite{Grebenkov06,Grebenkov19,Grebenkov20,Grebenkov22a}.
In particular, the statistics of encounters between a diffusing
particle and the confining boundary can be determined via a spectral
expansion based on the Dirichlet-to-Neumann operator
\cite{Grebenkov19c,Grebenkov21a,Grebenkov22d}.  Its
relations to other first-passage time statistics were also
investigated \cite{Grebenkov20,Grebenkov20c}.

The spectral properties of the Dirichlet-to-Neumann operator and the
related Steklov problem were thoroughly investigated for Euclidean
domains and Riemannian manifolds (see the recent book \cite{Levitin}
and reviews \cite{Girouard17,Colbois23}).  We focus here on the most
basic setting of an Euclidean domain $\Omega \subset \R^d$ with a
smooth bounded boundary $\pa$.  The Dirichlet-to-Neumann operator $\M$
is defined as a map of a function $f \in H^{\frac12}(\pa)$ on the
boundary $\pa$ onto another function $g \in H^{-\frac12}(\pa)$ on that
boundary such that $g = (\partial_n u)|_{\pa} = \M f$, where
$\partial_n$ is the normal derivative oriented outwards the domain
$\Omega$, and $u$ is the unique solution of the Dirichlet boundary
value problem:
\begin{equation}  \label{eq:Laplace}
\Delta u = 0 \quad \textrm{in}~\Omega,  \qquad  u = f  \quad \textrm{on}~\pa,
\end{equation}
with $\Delta$ being the Laplace operator (here $H^{\pm \frac12}(\pa)$
are appropriate functional spaces, see Ref. \cite{Levitin} for
mathematical details and references).  In other words, the operator
$\M$ transforms the Dirichlet boundary condition $u|_{\pa} = f$ on
$\pa$ into an equivalent Neumann boundary condition $(\partial_n
u)|_{\pa} = \M f$.  For instance, if $f$ is a given concentration or
temperature profile maintained on $\pa$, then the Laplace equation
describes the steady-state regime of molecular or heat diffusion, and
$(\partial_n u)|_{\pa}$ is proportional to the flux density on $\pa$.

When $\Omega$ is bounded, $\M$ is known to be pseudo-differential
self-adjoint operator with a discrete spectrum, i.e., there is an
infinite countable sequence of eigenpairs $\{\mu_k, v_k\}$ satisfying
$\M v_k = \mu_k v_k$; the nonnegative eigenvalues $\mu_k$ are
enumerated by $k = 0,1,2,\ldots$ in an increasing order,
\begin{equation}  \label{eq:muk_order}
0 \leq \mu_0 \leq \mu_1 \leq \ldots \nearrow +\infty, 
\end{equation}
whereas the associated eigenfunctions $\{v_k\}$ form a complete basis
in the space $L^2(\pa)$ of square-integrable functions on $\pa$
\cite{Levitin}.  Alternatively, one can search for solutions of the
Steklov problem,
\begin{equation}
\Delta V_k = 0   \quad \textrm{in}~\Omega,  \qquad  \partial_n V_k = \mu_k V_k  \quad \textrm{on}~\pa,
\end{equation}
where the Steklov eigenvalues standing in the boundary condition are
identical to $\mu_k$.  This tight relation implies that each Steklov
eigenfunction $V_k$ can be obtained as a harmonic extension of the
eigenfunction $v_k$ of $\M$.

Despite numerous mathematical studies of spectral properties of the
Dirichlet-to-Neumann operator \cite{Levitin}, intricate relations
between its spectrum and the geometric features of the boundary $\pa$
are not yet fully understood.  The eigenvalues and eigenfunctions of
$\M$ are known explicitly only in few simple domains such as a ball, a
space between concentric spheres, the exterior of a ball, and
rectangular cuboids \cite{Levitin,Grebenkov20b}.  In particular, the
role of the boundary anisotropy remains unclear.
The situation is even worse for the exterior problem when $\Omega =
\R^d \backslash \Omega_0$ is the exterior of a bounded domain
$\Omega_0$.  Even though the domain $\Omega$ is unbounded, its
boundary $\pa$ is bounded that implies again the discrete spectrum of
the Dirichlet-to-Neumann operator $\M$
\cite{Auchmuty13,Auchmuty14,Arendt15,Auchmuty18,Christiansen23,Xiong23}.
However, the analysis of the exterior problem is more difficult; in
particular, the mathematical proofs substantially differ for space
dimensions $d = 2$ and $d \geq 3$.  Relations between spectral
properties and the geometric shape of $\pa$ were much less studied.
For instance, to our knowledge, the exterior of a ball is the unique
example, for which the eigenvalues and eigenfunctions of $\M$ are
known explicitly for the exterior problem.

In this paper, we study the spectral properties of the
Dirichlet-to-Neumann operator $\M$ on prolate and oblate spheroidal
surfaces that allow one to model various anisotropic shapes in three
dimensions, ranging from a needle to a disk.  We focus on the less
studied exterior spectral problem (an extension to the interior
problem is summarized in Appendix \ref{sec:int}).  By employing the
prolate/oblate spheroidal coordinates to represent a general solution
of the Laplace equation (\ref{eq:Laplace}), we obtain a convenient
matrix representation of the operator $\M$.  This matrix can then be
truncated and diagonalized numerically to approximate the eigenvalues
$\mu_k$ and eigenfunctions $v_k$ of $\M$.  This efficient technique
allows us to investigate how the spectral properties of the
Dirichlet-to-Neumann operator depend on the anisotropy of the
boundary, especially in the limits of elongated (needle-like) and
flattened (disk-like) spheroids.  While similar techniques were
applied in the past for solving various boundary value problems in
spheroidal domains (see, e.g.,
\cite{Morse,Smythe,Barucq10,LeGia11,Costea11,Luo15,Gomez15,Xue17,Traytak18,Piazza19,Chaigneau22,Chaigneau23}
and references therein), we are not aware of earlier studies of the
spectral properties of the Dirichlet-to-Neumann operator in these
domains.  Sections \ref{sec:prolate} and \ref{sec:oblate} are devoted
respectively to prolate and oblate spheroidal domains.  In
Sec. \ref{sec:application}, we discuss two applications of these
results for understanding diffusion-controlled reactions and the
statistics of boundary encounters.  Section \ref{sec:conclusion}
summarizes our findings and presents future perspectives.

\section{Prolate spheroids}
\label{sec:prolate}

\begin{figure}
\centering
\includegraphics[width=88mm]{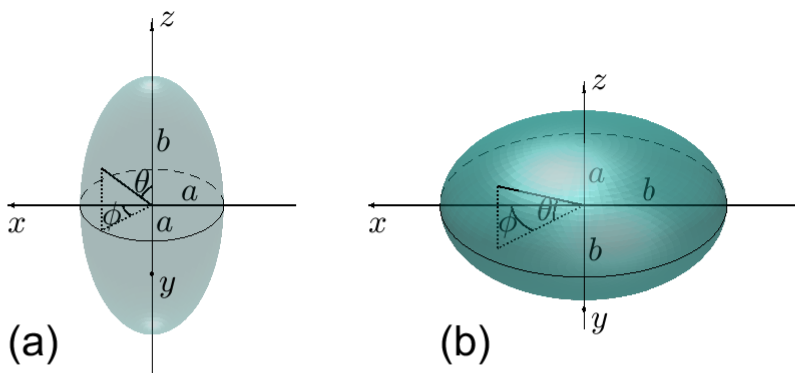} 
\caption{
Illustration for prolate {\bf (a)} and oblate {\bf (b)} spheroidal
coordinates $(\alpha,\theta,\phi)$.  Note that the angle $\theta$ is
defined differently in two cases. }
\label{fig:scheme}
\end{figure}

In this section, we study the Dirichlet-to-Neumann operator $\M$ in
the exterior of a prolate spheroid with semi-axes $a \leq b$:
\begin{equation}
\Omega = \biggl\{ (x,y,z)\in \R^3 ~:~ \frac{x^2}{a^2} + \frac{y^2}{a^2} + \frac{z^2}{b^2} > 1 \biggr\} .
\end{equation}  
In the prolate spheroidal coordinates $(\alpha,\theta,\phi)$, 
\begin{equation*}
\left(\begin{array}{c} x \\ y \\ z \\ \end{array}\right) = a_E \left(\begin{array}{c} 
\sinh \alpha \sin \theta \cos\phi \\ 
\sinh \alpha \sin \theta \sin\phi \\ 
\cosh \alpha \cos \theta  \\  \end{array} \right)   \qquad
\left\{\begin{array}{l} 0 < \alpha < \infty \\ 0 \leq \theta \leq \pi \\ 0 \leq \phi < 2\pi \\ \end{array} \right\},
\end{equation*}
with $a_E = \sqrt{b^2 - a^2}$ (Fig. \ref{fig:scheme}a), the
scale factors determining the surface and volume elements, are
\cite{Korn}
\begin{subequations}  \label{eq:h_prolate}
\begin{align}
h_\alpha & = h_\theta = a_E \sqrt{\sinh^2\alpha + \sin^2 \theta}, \\
h_\phi & = a_E \sinh\alpha \sin\theta .
\end{align}
\end{subequations}
In these coordinates, the domain $\Omega$ is characterized by $\alpha
> \alpha_0 = \tanh^{-1}(a/b)$, while its boundary $\pa$ is determined
by the condition $\alpha = \alpha_0$.  The action of the Laplace
operator onto a function $u$ reads
\begin{align} \nonumber
\Delta u & = \frac{1}{a_E^2 (\sinh^2\alpha + \sin^2\theta)} \biggl[\frac{1}{\sinh \alpha}
\frac{\partial}{\partial\alpha} \biggl(\sinh\alpha \frac{\partial u}{\partial \alpha}\biggr) \\  \label{eq:Laplace_prolate}
& + \frac{1}{\sin\theta} \frac{\partial}{\partial\theta} \biggl(\sin\theta \frac{\partial u}{\partial \theta}\biggr) \biggr] 
+ \frac{1}{a_E^2 \sinh^2\alpha \, \sin^2\theta} \frac{\partial^2 u}{\partial \phi^2} \,.
\end{align}

\subsection*{Matrix representation}

A general solution of the Laplace equation reads
\begin{equation}  \label{eq:u_prolate}
u(\alpha,\theta,\phi) = \sum\limits_{n=0}^\infty \sum\limits_{m=-n}^n A_{mn} Q_n^m(\cosh \alpha) Y_{mn}(\theta,\phi) ,
\end{equation}
where $A_{mn}$ are arbitrary coefficients, $Q_n^m(z)$ are the
associated Legendre functions of the second kind (see Appendix
\ref{sec:numerics}), and
\begin{align}
Y_{mn}(\theta,\phi) & = a_{mn} P_n^m(\cos\theta) e^{im\phi}  , \\  \nonumber
a_{mn} & = \sqrt{\frac{2n+1}{4\pi} \, \frac{(n-m)!}{(n+m)!}}
\end{align}
are the normalized spherical harmonics, with $P_n^m(x)$ being the
associated Legendre polynomials.  The normal derivative reads
\begin{align} 
& \left. \frac{\partial u}{\partial n} \right|_{\pa} 
= - \left. \frac{1}{h_\alpha} \partial_\alpha u  \right|_{\alpha=\alpha_0}\\  \nonumber
& = - \sum\limits_{n=0}^\infty \sum\limits_{m=-n}^n A_{mn} Y_{mn}(\theta,\phi) 
\frac{\sinh \alpha_0 \, Q^{'m}_n(\cosh \alpha_0)}{a_E \sqrt{\cosh^2\alpha_0 - \cos^2\theta}} \,,
\end{align}
where prime denotes the derivative with respect to the argument.  In
other words, for a square-integrable function $f$ on $\pa$, decomposed
on the complete basis of spherical harmonics,
\begin{equation}  \label{eq:f_Ymn}
f(\theta,\phi) = \sum\limits_{n=0}^\infty \sum\limits_{m=-n}^n  f_{mn} Y_{mn}(\theta,\phi)
\end{equation}
(with coefficients $f_{mn} = A_{mn} Q_n^m(\cosh\alpha_0)$), the action
of the Dirichlet-to-Neumann operator $\M$ is
\begin{equation}  \label{eq:auxil1a}
[\M f](\theta,\phi) = \sum\limits_{n=0}^\infty \sum\limits_{m=-n}^n  f_{mn}  \frac{c_{mn} \, Y_{mn}(\theta,\phi)}
{\sqrt{\cosh^2\alpha_0 - \cos^2\theta} } \,,
\end{equation}
where
\begin{equation}  \label{eq:cmn_prolate}
c_{mn} = - \frac{\sinh \alpha_0 \, Q^{'m}_n(\cosh \alpha_0)}{a_E Q_n^m(\cosh\alpha_0)} \,.
\end{equation}
At the same time, the function $\M f$ can also be decomposed on the
complete basis of $Y_{mn}$:  
\begin{equation}  \label{eq:Mf_gmn}
[\M f](\theta,\phi) = \sum\limits_{n=0}^\infty \sum\limits_{m=-n}^n  g_{mn}  Y_{mn}(\theta,\phi).
\end{equation}
Its coefficients $g_{mn}$ can be obtained by multiplying
Eq. (\ref{eq:auxil1a}) by $Y_{m'n'}^*(\theta,\phi) \sin \theta$ and
integrating over $\theta$ and $\phi$:
\begin{align*}
g_{m'n'} & = \int\limits_0^{2\pi} d\phi \int\limits_0^{\pi} d\theta \, \sin\theta \, Y_{m'n'}^*(\theta,\phi) \\ 
& \times \sum\limits_{n=0}^\infty \sum\limits_{m=-n}^n  f_{mn}  \frac{c_{mn} \, Y_{mn}(\theta,\phi)}
{\sqrt{\cosh^2\alpha_0 - \cos^2\theta} } \\
& = \sum\limits_{n=0}^\infty \sum\limits_{m=-n}^n  f_{mn} \MM_{mn,m'n'}  ,
\end{align*}
where
\begin{equation}  \label{eq:M_prolate}  
\MM_{mn,m'n'} = 2\pi \, \delta_{m,m'} a_{mn} a_{mn'}  \, c_{m'n'} \, F_{n,n'}^m(\cosh\alpha_0),
\end{equation}
and
\begin{equation}   \label{eq:Fmnn}
F_{n,n'}^m(z) = \int\limits_{-1}^1 dx \frac{P_n^m(x) P_{n'}^m(x)}{\sqrt{z^2 - x^2} }  \,.
\end{equation}
We stress that the elements of the (infinite-dimensional) matrix $\MM$
are enumerated by double indices $mn$ and $m'n'$.  In practice, we
employ the following order:
\begin{equation*}
\begin{array}{c  c   c c c c c c c c    c}
mn: & 00 & (-1)1 & 01 & 11 & (-2)2 & (-1)2 & 02 & 12 &  22 & \ldots \\ \end{array}
\end{equation*} 
which is borrowed from the enumeration of spherical harmonics.  As a
consequence, Eq. (\ref{eq:Mf_gmn}) can be written as
\begin{equation}
[\M f](\theta,\phi) = \sum\limits_{n=0}^\infty \sum\limits_{m=-n}^n  [\MM f]_{mn}  Y_{mn}(\theta,\phi) ,
\end{equation}
i.e., the matrix $\MM$ represents the operator $\M$ on the orthonormal
basis of spherical harmonics.

The eigenvalues of the matrix $\MM$ coincide with the eigenvalues
$\mu_k$ of the Dirichlet-to-Neumann operator $\M$; in turn, each
eigenvector $\VV_k$ of $\MM$, satisfying $\MM \VV_k = \mu_k \VV_k$,
determines the coefficients in the representation of the associated
eigenfunction $v_k$ on the basis of spherical harmonics:
\begin{equation}  \label{eq:vk_prolate}
v_k(\theta,\phi) = \sum\limits_{n=0}^\infty \sum\limits_{m=-n}^n [\VV_k]_{mn} Y_{mn}(\theta,\phi).
\end{equation} 
In Appendix \ref{sec:orthogonality}, we check the orthogonality of
these eigenfunctions to each other; moreover, they can also be
normalized as
\begin{align}  \nonumber
& \bigl(v_k, v_{k'}\bigr)_{L^2(\pa)} = \int\limits_{\pa} d{\bf s}\, v_k^* \,v_{k'} \\   \label{eq:vk_orthogonal}
& \quad = \int\limits_0^{\pi} d\theta \int\limits_0^{2\pi} d\phi  \, h_\theta \, h_\phi \, v_k^*(\theta,\phi) 
\, v_{k'}(\theta,\phi) = \delta_{k,k'} .
\end{align}
Using Eq. (\ref{eq:u_prolate}) for a general solution of the Laplace
operator, one easily finds a harmonic extension of the eigenfunction
$v_k$ into $\Omega$, i.e., the Steklov eigenfunction associated to the
eigenvalue $\mu_k$:
\begin{equation}  \label{eq:Vk_prolate}
V_k(\alpha,\theta,\phi) = \sum\limits_{n=0}^\infty \sum\limits_{m=-n}^n  
\frac{Q_n^m(\cosh \alpha)}{Q_n^m(\cosh \alpha_0)} [\VV_k]_{mn} Y_{mn}(\theta,\phi).
\end{equation} 
The asymptotic behavior of $Q_n^m(z)$ for large $|z|$, $Q_n^m(z)
\propto z^{-n-1}$ implies the expected power-law decay of
Steklov eigenfunctions in the leading order as $|\x|\to \infty$:
\begin{align}  \nonumber
V_k(\alpha,\theta,\phi) & \simeq  \frac{[\VV_k]_{00}}{\sqrt{4\pi}\, Q_0(\cosh\alpha_0)} \, \frac{1}{\cosh\alpha} \\
& \simeq  \frac{[\VV_k]_{00}}{\sqrt{4\pi}\, Q_0(\cosh\alpha_0)} \frac{a_E}{|\x|} \propto |\x|^{-1}  ,
\end{align}
where $|\x| = a_E \sqrt{\cosh^2\alpha - \sin^2\theta}$ is the distance
from the origin to a point $\x = (\alpha,\theta,\phi)$.

\subsection*{Classification of eigenfunctions}

Since the Dirichlet-to-Neumann operator $\M$ does not affect the angle
$\phi$, the matrix elements $\MM_{mn,m'n'}$ are nonzero only when $m =
m'$.  In other words, the action of $\M$ onto a function $f(\theta)
e^{im\phi}$ does not alter its dependence on $\phi$: $\M (f(\theta)
e^{im\phi}) = g(\theta) e^{im\phi}$.  As a consequence, any
eigenfunction $v_k$ depends on $\phi$ via a factor $e^{im\phi}$ for
some integer $m$ (or via a linear combination of $e^{im\phi}$ and
$e^{-im\phi}$, see below).
This property allows one to classify all eigenfunctions according to
their dependence on $\phi$ and thus to enumerate them as
$v_{mn}(\theta,\phi)$, in analogy to spherical harmonics
$Y_{mn}(\theta,\phi)$.  Here the index $m$ determines the dependence
of the eigenfunction on $\phi$, $v_{mn}(\theta,\phi) \propto
e^{im\phi}$, whereas the nonnegative index $n =
|m|,|m|+1,|m|+2,\ldots$ enumerates all such functions so that the
associated eigenvalues $\mu_{mn}$ appear in an increasing order (for
each fixed $m$):
\begin{equation}  \label{eq:mu_mn_order}
0 \leq \mu_{m|m|} \leq \mu_{m(|m|+1)} \leq \ldots 
\end{equation}
Note that the index $n$ starts from $|m|$ in order to automatically
satisfy the conventional restriction $|m| \leq n$, known for spherical
harmonics.

An alternative way to look at this classification consists in
representing the matrix $\MM$ as
\begin{equation}  \label{eq:M_sum}
\MM = \sum\limits_{m=-\infty}^\infty \MM_m ,
\end{equation}
where the (sub)matrix $\MM_m$ is composed of elements $\MM_{mn,mn'}$
(and $0$ otherwise).  For instance, one has%
\footnote{
Note that $\MM_{0n,0n'} = 0$ for odd $n+n'$ due to the symmetry
(\ref{eq:ALegendre_symm}) of the associated Legendre polynomials and
the fact that the function $1/\sqrt{z^2-x^2}$ in Eq. (\ref{eq:Fmnn})
is symmetric.}
%
\begin{equation} \label{eq:M0}
{\tiny
\MM_0 = \left(\begin{array}{c | c c c | c c c c c | c}
\MM_{00,00} & 0 & 0 & 0 & 0 & 0 & \MM_{00,02} & 0 & 0 & \ldots \\   \hline
    0       & 0 & 0 & 0 & 0 & 0 & 0 & 0 & 0 & \ldots \\ 
    0       & 0 & \MM_{01,01} & 0 & 0 & 0 & 0 & 0 & 0 & \ldots \\ 
    0       & 0 & 0 & 0 & 0 & 0 & 0 & 0 & 0 & \ldots \\  \hline
    0       & 0 & 0 & 0 & 0 & 0 & 0 & 0 & 0 & \ldots \\ 
    0       & 0 & 0 & 0 & 0 & 0 & 0 & 0 & 0 & \ldots \\ 
\MM_{02,00} & 0 & 0 & 0 & 0 & 0 & \MM_{02,02} & 0 & 0 & \ldots \\ 
    0       & 0 & 0 & 0 & 0 & 0 & 0 & 0 & 0 & \ldots \\ 
    0       & 0 & 0 & 0 & 0 & 0 & 0 & 0 & 0 & \ldots \\ \hline
\ldots & \ldots & \ldots & \ldots & \ldots & \ldots & \ldots & \ldots & \ldots & \ldots  \\ 
\end{array}  \right).}
\end{equation}
If a vector $\XX$ has the form $\XX_{mn} = \delta_{m,m_0} x_n$ (i.e.,
its nonzero elements appear only for $m = m_0$), then $\MM_m \XX = 0$
for all $m \ne m_0$.  In other words, the space $\ell^2$ of all
vectors with square-summable elements can be decomposed into an
(infinite) direct product of subspaces $\ell^2_m$ enumerated by $m$
ranging from $-\infty$ to $+\infty$.  As a consequence, one can
diagonalize separately each matrix $\MM_m$ and then combine their
eigenvalues $\mu_{mn}$ and eigenvectors $\VV_{mn}$ (enumerated by the
index $n = |m|,|m|+1,\ldots$) to construct the eigenvalues and
eigenvectors of the matrix $\MM$.  We expect that the union of all
eigenvalues $\mu_{mn}$ gives {\it all} eigenvalues of the matrix
$\MM$, i.e., such a decomposition determines the whole spectrum of
$\MM$.  This statement is elementary in the finite-dimensional case,
in particular, for a truncation of the matrix $\MM$ that we will use
for numerical computations.  We can then rewrite the expansions
(\ref{eq:vk_prolate}, \ref{eq:Vk_prolate}) as
\begin{subequations}
\begin{align}  \label{eq:vmn_prolate}
v_{mn}(\theta,\phi) & = \sum\limits_{n'=|m|}^\infty [\VV_{mn}]_{mn'} Y_{mn'}(\theta,\phi), \\  \label{eq:Vmn_prolate}
V_{mn}(\alpha,\theta,\phi) & = \sum\limits_{n'=|m|}^\infty \frac{Q_{n'}^m(\cosh \alpha)}{Q_{n'}^m(\cosh\alpha_0)} 
[\VV_{mn}]_{mn'} Y_{mn'}(\theta,\phi). 
\end{align}
\end{subequations}

We note that the symmetry
\begin{equation} \label{eq:ALegendre_symm}
P_n^m(-x) = (-1)^{m+n} P_n^m(x), 
\end{equation}
implies $\MM_{-m} = \MM_m$ that allows one to restrict $m$ to be
nonnegative.  As a consequence, any eigenvalue $\mu_{mn}$ with $m\ne
0$ should be (at least) twice degenerate.  This degeneracy implies
that any linear combination of $v_{(-m)n}$ and $v_{mn}$ is also an
eigenfunction.  As a consequence, if the eigenfunctions are
constructed by using the decomposition (\ref{eq:vk_prolate}) based on
the diagonalization of the whole matrix $\MM$ and then classified
according to their dependence on $\phi$, each eigenfunction $v_k$ may
in general exhibit the dependence on $\phi$ as a linear combination of
$e^{im\phi}$ and $e^{-im\phi}$.  An appropriate rotation by a $2\times
2$ matrix can transform a pair of such eigenfunctions into those that
are proportional to $e^{im\phi}$ and $e^{-im\phi}$.  However, this
step is not needed in practice, as we will diagonalize the matrices
$\MM_m$ to produce directly the desired dependence $e^{im\theta}$ on
$\phi$.

In the following, we use interchangeably both notations $\mu_k, v_k$
and $\mu_{mn}, v_{mn}$ for eigenvalues and eigenfunctions.  We recall
that the single-index enumeration relies on the global ordering of all
eigenvalues $\mu_k$ in Eq.  (\ref{eq:muk_order}).  In turn, the
double-index enumeration is based on the symmetries of eigenfunctions,
namely, on their dependence on $\phi$ via $e^{im\phi}$, whereas the
second index $n$ employs the ordering of $\mu_{mn}$ for each $m$ in
Eq. (\ref{eq:mu_mn_order}).

\subsection*{Limit of a sphere}

In the limit $a \to b$, the prolate spheroid approaches the sphere of
radius $b$.  In this limit, one has $a_E \to 0$ and $\alpha_0 \to
\infty$ such that $a_E \cosh \alpha_0 = b$ remains constant.  As a
consequence, the coefficients in Eq. (\ref{eq:cmn_prolate}) diverge as
$c_{mn} \approx \tfrac{n+1}{b} \, \sinh \alpha_0$, whereas the
matrix elements in Eq. (\ref{eq:Fmnn}) behave in the leading order as
\begin{equation}  \label{eq:Fnnm_sphere}
F_{n,n'}^m(\cosh\alpha_0) \approx \frac{\delta_{n,n'}}{2\pi a_{mn}^2 \cosh \alpha_0} \,,
\end{equation} 
implying that $\MM_{mn,m'n'} \to \tfrac{n+1}{b} \delta_{n,n'}
\delta_{m,m'}$.  The diagonal structure of this matrix yields
\begin{equation}
\mu_{mn} = \frac{n+1}{b} \,, \qquad  v_{mn} = \frac{1}{b} \, Y_{mn}(\theta,\phi).  
\end{equation}
We retrieve therefore the well-known eigenvalues and eigenfunctions of
the Dirichlet-to-Neumann operator for the exterior of a sphere.  Note
that the eigenvalues do not depend on $m$; moreover, since $m$ ranges
from $-n$ to $n$, the degeneracy of the eigenvalue $\mu_{mn}$ is
$2n+1$, as expected for a sphere due to its rotational symmetry.  As
illustrated below, the anisotropy of spheroids breaks this symmetry
and reduces the degeneracy of eigenvalues.

\subsection*{Numerical implementation}

For a practical implementation, the infinite-dimensional matrix $\MM$
has to be truncated to a finite size.  In a basic setup, one can
choose the truncation order $\nmax$ to keep $n = 0,1,2,\ldots,\nmax$,
and then construct the truncated matrix of size $(\nmax+1)^2
\times (\nmax+1)^2$, as detailed in Appendix \ref{sec:numerics}.  A
numerical diagonalization of the truncated matrix provides an
approximation for a number of eigenvalues and eigenfunctions of $\M$.
As illustrated below, the accuracy of this approximation increases
rapidly with the truncation order $\nmax$.

A much faster procedure consists in dealing with the reduced matrices
$\hat{\MM}_m$, which are obtained from $\MM_m$ by removing zero
columns and rows.  For instance, the matrix $\MM_0$ from
Eq. (\ref{eq:M0}) has the following reduced form
\begin{equation}   \label{eq:hatM0}
\hat{\MM}_0 = \left(\begin{array}{c c c c} 
\MM_{00,00} &       0     & \MM_{00,02} & \ldots \\
     0      & \MM_{01,01} &      0      & \ldots \\
\MM_{02,00} &       0     & \MM_{02,02} & \ldots \\
\ldots & \ldots & \ldots & \ldots \\ \end{array} \right).
\end{equation}
In practice, we start from the truncated matrix $\MM$ of size
$(\nmax+1)^2 \times (\nmax+1)^2$ and then dispatch its columns and
rows according to $m$ ranging from $0$ to $\nmax$, into matrices
$\hat{\MM}_m$.  As the reduced matrix $\hat{\MM}_m$ is of much smaller
size $(\nmax+1-m) \times (\nmax+1-m)$, its numerical diagonalization
is significantly faster.  Its eigenvalues approximate the eigenvalues
$\mu_{mn}$ of the Dirichlet-to-Neumann operator $\M$; in turn, its
eigenvectors determine the eigenfunctions $v_{mn}$ of $\M$ via
Eq. (\ref{eq:vmn_prolate}), and the Steklov eigenfunctions $V_{mn}$
via Eq. (\ref{eq:Vmn_prolate}).  The former eigenfunctions are
orthogonal to each other by construction.  In turn, one needs to
impose their normalization according to Eq. (\ref{eq:vk_orthogonal}).
We recall that this normalization is fixed up to an arbitrary phase
factor $e^{i\alpha}$.  Further simplifications can be achieved for the
axisymmetric problem, see Appendix \ref{sec:axi}.
It is worth noting that the reduced matrices $\hat{\MM}_m$ form a
block-diagonal matrix
\begin{equation}   \label{eq:hatM}
\hat{\MM} = \left(\begin{array}{c c c c c} 
\ldots &    \ldots      &   \ldots    & \ldots      & \ldots\\
\ldots & \hat{\MM}_{-1} &      0      &       0     & \ldots\\
\ldots &      0         & \hat{\MM}_0 &       0     & \ldots \\
\ldots &      0         &      0      & \hat{\MM}_1 & \ldots \\
\ldots &    \ldots      &    \ldots   &   \ldots    & \ldots \\ \end{array} \right),
\end{equation}
which can be seen as a transformation of the original matrix $\MM$ by
re-ordering its columns and rows.  Since these two matrices have the
same eigenvalues, it is much faster to search for the eigenvalues of
the matrix $\hat{\MM}$, which are obtained by combining the
eigenvalues of its diagonal blocks $\hat{\MM}_m$.

\begin{figure}
\begin{center}
\includegraphics[width=88mm]{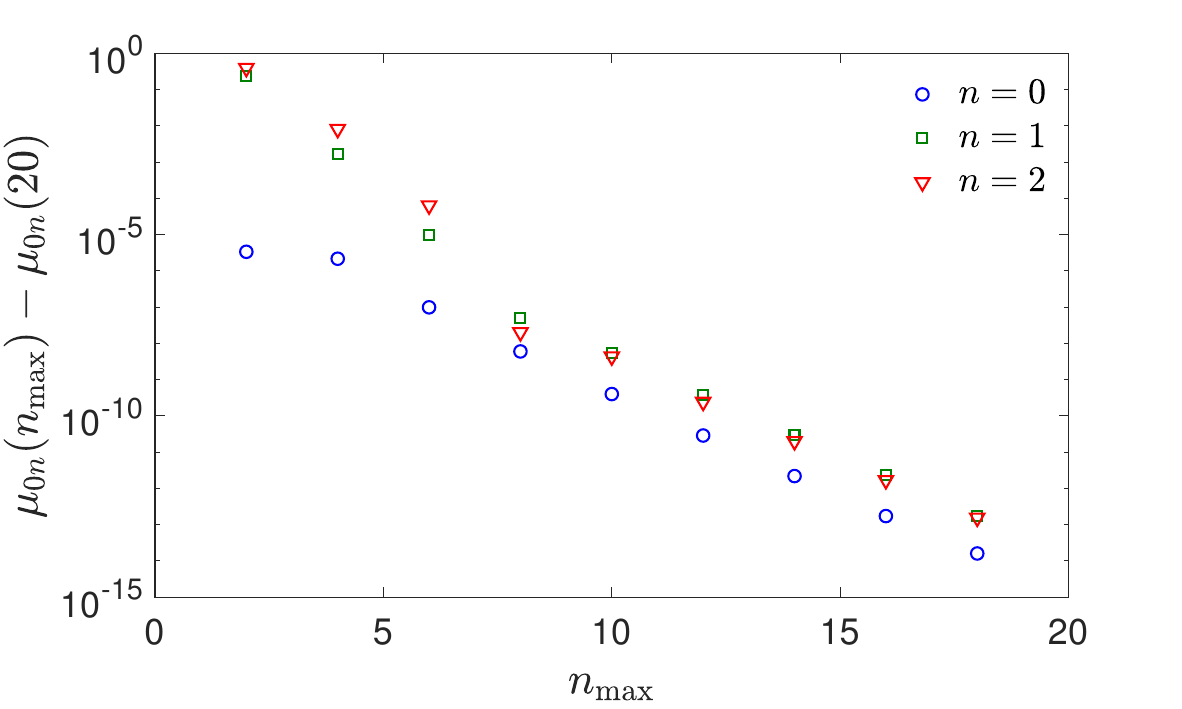} 
\end{center}
\caption{
Illustration for the convergence of the numerical method in the case
of the exterior of the prolate spheroid with semi-axes $a = 0.5$ and
$b = 1$.  Three eigenvalues $\mu_{00}(\nmax)$, $\mu_{01}(\nmax)$, and
$\mu_{02}(\nmax)$ of the truncated matrix $\MM$ of size $(\nmax+1)^2
\times (\nmax+1)^2$ as functions of the truncation order $\nmax$ are
shown by symbols.  Their values at $\nmax = 20$, considered here as a
benchmark, are subtracted to estimate the error of truncation.  }
\label{fig:mu00_accuracy}
\end{figure}

To illustrate the fast convergence of the numerical method, we choose
the prolate spheroid with semi-axes $a = 0.5$ and $b = 1$ and compute
three eigenvalues $\mu_{00}(\nmax)$, $\mu_{01}(\nmax)$, and
$\mu_{02}(\nmax)$ of the truncated matrix $\MM$ of size $(\nmax+1)^2
\times (\nmax+1)^2$ as functions of the truncation order $\nmax$.
To estimate the truncation error, we subtract from $\mu_{0n}(\nmax)$
its value at $\nmax = 20$, considered as a proxy of the limiting value
$\mu_{0n}(\infty)$.  Figure \ref{fig:mu00_accuracy} shows how fast the
error decreases with $\nmax$.  We checked the rapid convergence for
other values of $m$ and $n$, as well as for various aspect ratios (not
shown).  In particular, we found that larger truncation orders may be
needed to achieve high accuracy when $a/b$ gets smaller.  Moreover,
when $a/b$ is close to $1$, the matrix elements
$F_{n,n'}^m(\cosh\alpha_0)$ vanish according to
Eq. (\ref{eq:Fnnm_sphere}) that requires a suitable rescaling and
further improvements of the numerical procedure described in Appendix
\ref{sec:numerics}.  Since the case of a sphere is fully explicit, we
do not study geometric settings of almost spherical domains.

A similar error analysis was performed for other problems (interior of
a prolate spheroid, oblate spheroids, see below).  In the following,
we generally use $\nmax = 10$ that is sufficient to produce accurate
numerical results for all the considered settings.

\subsection*{Examples of eigenfunctions}

\begin{figure}
\begin{center}
\includegraphics[width=88mm]{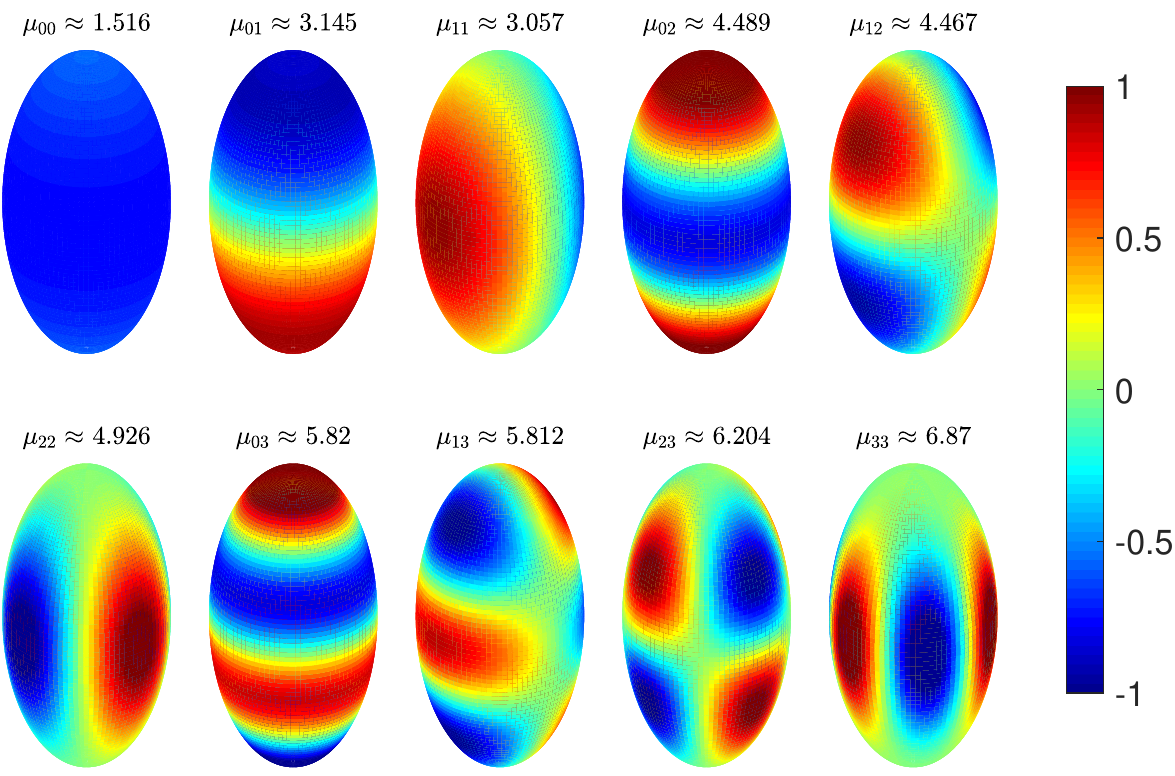} 
\end{center}
\caption{
Several eigenfunctions $v_{mn}$ of the Dirichlet-to-Neumann operator
$\M$ for the exterior of the prolate spheroid with semi-axes $a = 0.5$
and $b = 1$.  The associated eigenvalues are shown on the top.  The
eigenfunctions with even $m+n$ are symmetric with respect to the
horizontal plane $z = 0$, whereas the eigenfunctions with odd $m+n$
are antisymmetric, in agreement with Eq. (\ref{eq:vmn_symmetry}).  The
truncation order is $\nmax = 10$.  }
\label{fig:vn_prolateExt}
\end{figure}

Figure \ref{fig:vn_prolateExt} illustrates ten eigenfunctions $v_{mn}$
of the Dirichlet-to-Neumann operator $\M$ for the exterior of the
prolate spheroid with semi-axes $a = 0.5$ and $b = 1$.  The ground
eigenfunction $v_{00}$ is not constant (see below), even though its
minor changes are difficult to see due to the chosen colorbar, for
which color changes in the same range of values from $-1$ to $1$ for
all shown eigenfunctions.  The geometric structure of the remaining
shown eigenfunctions resembles that of the spherical harmonics
$Y_{mn}(\theta,\phi)$.  Note that the eigenfunctions $v_{mn}$ and
$v_{(-m)n}$ correspond to the same eigenvalue $\mu_{mn}$ and differ
only by the factor $e^{\pm im\phi}$; for this reason, the
eigenfunctions $v_{(-m)n}$ are not shown.

\begin{figure}
\begin{center}
\includegraphics[width=88mm]{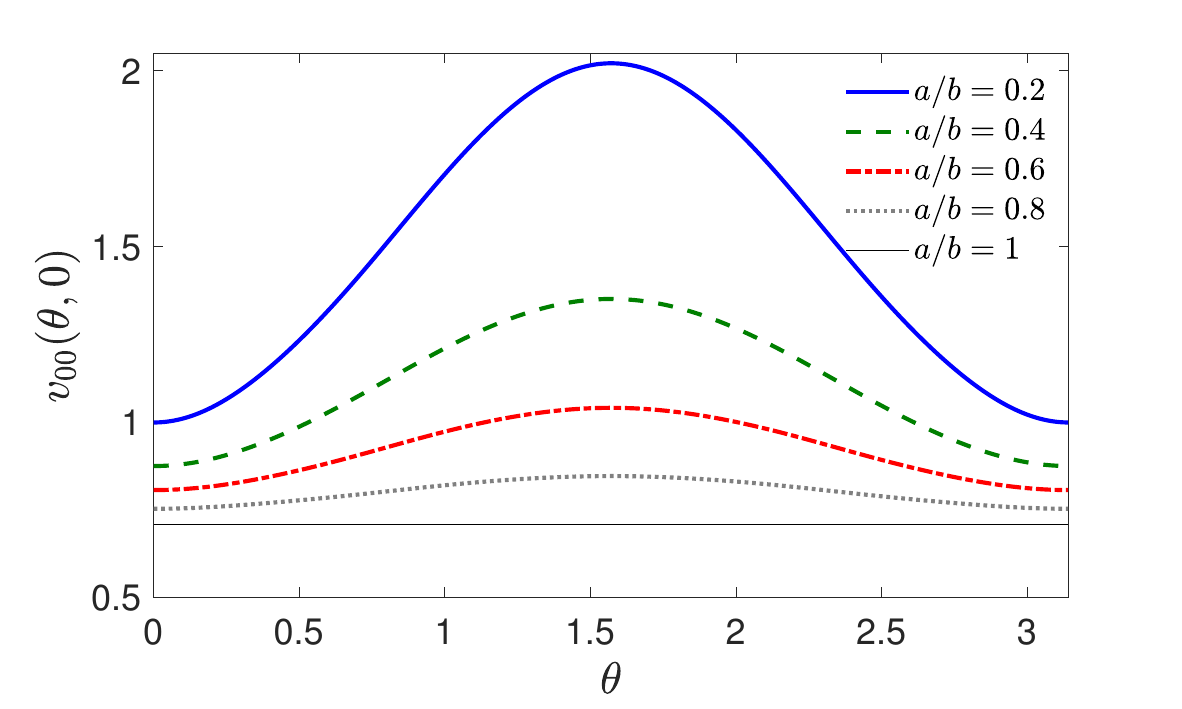} 
\includegraphics[width=88mm]{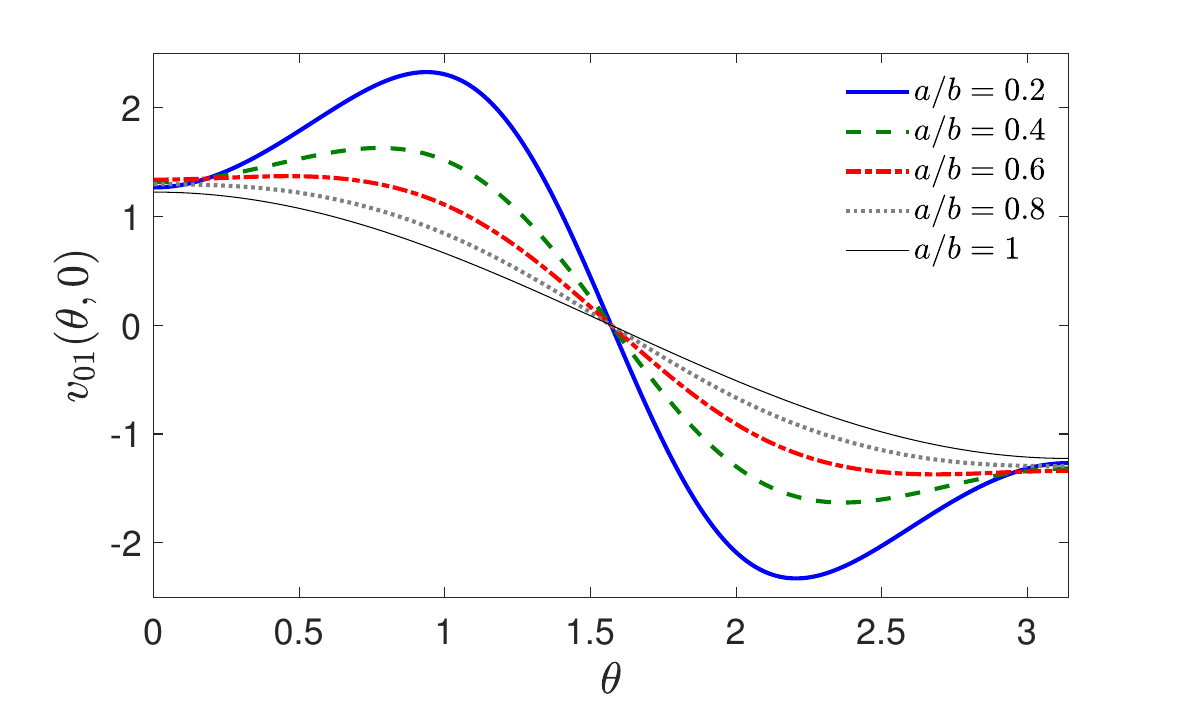} 
\includegraphics[width=88mm]{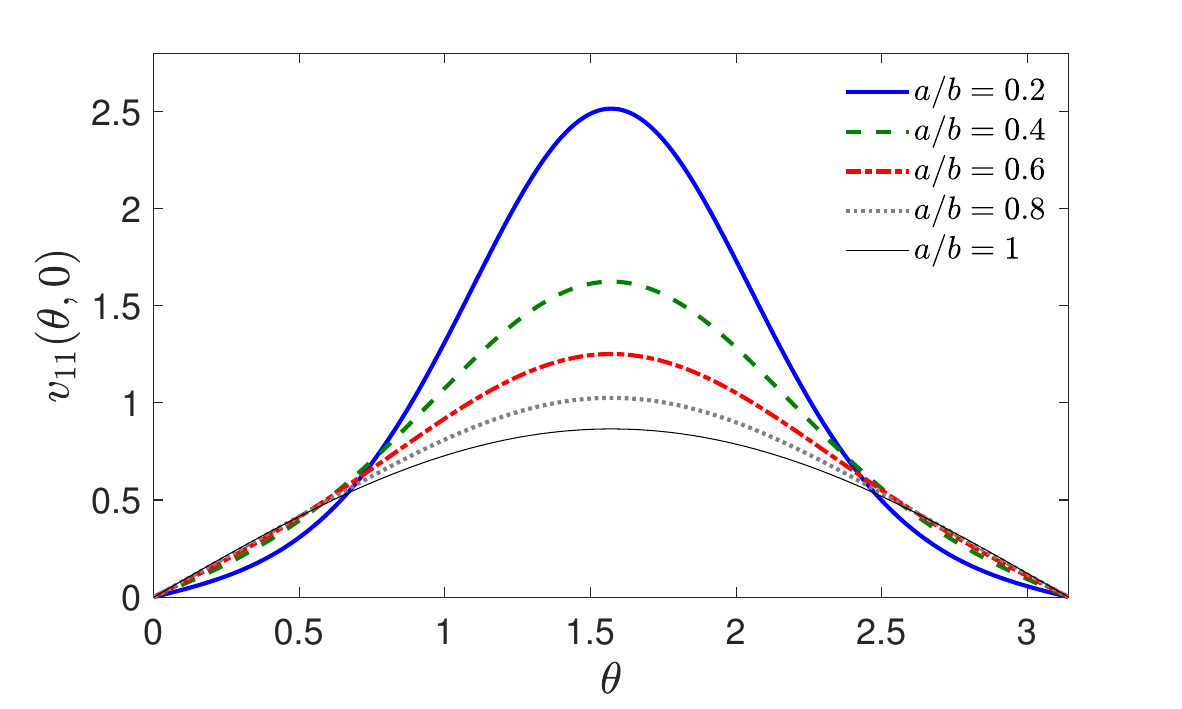} 
\end{center}
\caption{
The $xz$ projection of the eigenfunctions $v_{mn}(\theta,\phi)$ of the
Dirichlet-to-Neumann operator $\M$ for the exterior of the prolate
spheroid with semi-axis $a$ and $b = 1$, with $mn = 00$ (top) $mn =
01$ (middle), and $mn = 11$ (bottom).  Thin line shows the normalized
spherical harmonics $Y_{mn}(\theta,0)$ that corresponds to $a/b = 1$.
The truncation order is $\nmax = 10$.}
\label{fig:v0_prolateExt}
\end{figure}

Figure \ref{fig:v0_prolateExt} presents the behavior of
$v_{mn}(\theta,\phi)$ as a function of $\theta$ for $\phi = 0$ (i.e.,
its projection onto the $xz$ plane) for prolate spheroidal surfaces of
variables minor semi-axis $a$ (with fixed $b$).  As $a\to b$, the
surface becomes spherical, and the eigenfunctions $v_{mn}$ coincide
with normalized spherical harmonics $Y_{mn}$.  In turn, as $a$
decreases, the eigenfunctions $v_{mn}$ deviate further and further
from spherical harmonics.  Note that $v_{00}$ and $v_{01}$ are
axisymmetric (independent of $\phi$), so that their structure, shown
for $\phi = 0$, remains the same for any $\phi$.  In turn, the
structure of $v_{11}$ is affected by the factor $e^{i\phi}$; in
particular, it is complex-valued, and its real part becomes negative
for $\phi$ between $\pi/2$ and $3\pi/2$.  Table
\ref{tab:mu_prolateExt} summarizes the first eigenvalues $\mu_{mn}$
for various prolate spheroids.

\begin{table}
\begin{tabular}{|c|c|c|c|c|c|c|}  \hline
$a/b$ & $\mu_{00}$ & $\mu_{01}$ & $\mu_{02}$ & $\mu_{11}$ & $\mu_{12}$ & $\mu_{22}$ \\ \hline
0.1  &  3.960  &  6.787  &  8.740  & 11.041  & 12.764  & 20.748 \\
0.2  &  2.558  &  4.820  &  6.523  &  6.035  &  7.649  & 10.797 \\
0.3  &  2.019  &  4.011  &  5.577  &  4.379  &  5.924  &  7.512 \\
0.4  &  1.717  &  3.514  &  4.964  &  3.554  &  5.034  &  5.889 \\
0.5  &  1.516  &  3.145  &  4.489  &  3.057  &  4.467  &  4.926 \\
0.6  &  1.367  &  2.844  &  4.092  &  2.721  &  4.054  &  4.287 \\
0.7  &  1.250  &  2.588  &  3.754  &  2.475  &  3.726  &  3.832 \\
0.8  &  1.153  &  2.365  &  3.464  &  2.285  &  3.451  &  3.490 \\
0.9  &  1.071  &  2.171  &  3.215  &  2.130  &  3.212  &  3.220 \\  
1.0  &  1      &  2      &  3      &  2      &  3      &   3    \\ \hline
\end{tabular}
\caption{
Several eigenvalues $\mu_{mn}$ of the Dirichlet-to-Neumann operator
$\M$ for the exterior of the prolate spheroid with semi-axes $a$ and
$b = 1$.  The matrices $\hat{\MM}_m$ are truncated to the size $(\nmax
+1 - m)\times (\nmax+1 - m)$, with $\nmax = 10$.  Further increase of
$\nmax$ did not change the shown eigenvalues. }
\label{tab:mu_prolateExt}
\end{table}

\subsection*{Asymptotic behavior for elongated spheroids}

\begin{figure}
\begin{center}
\includegraphics[width=88mm]{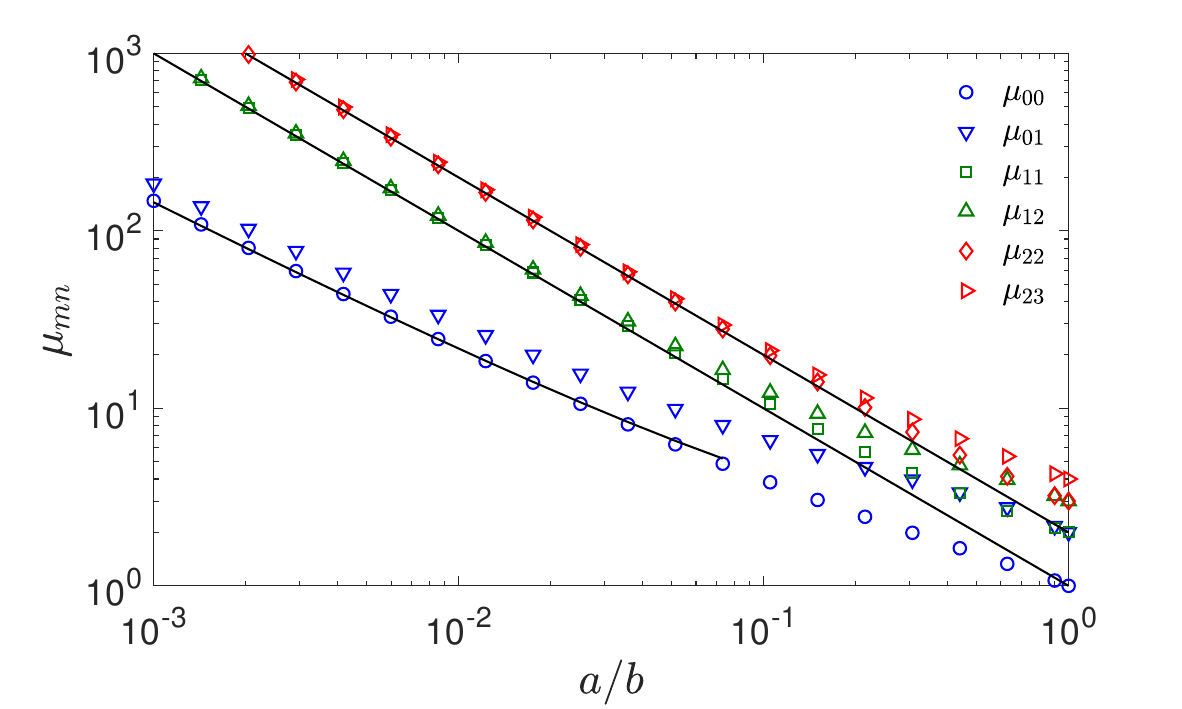} 
\end{center}
\caption{
Several eigenvalues $\mu_{mn}$ of the Dirichlet-to-Neumann operator
$\M$ for the exterior of the prolate spheroid as functions of its
minor semi-axis $a$, with $b = 1$.  At $a = b$, one retrieves the
eigenvalues $(n+1)/b$ for the exterior of a sphere of radius $b$.
Thin lines from bottom to top present $1/(a\ln(b/a))$, $1/a$ and $2/a$
that capture the asymptotic behavior of $\mu_{mn}$ with $m=0,1,2$,
respectively.  The eigenvalues are obtained by diagonalizing
$\hat{\MM}_m$ truncated to the size $(\nmax+1-m)\times (\nmax+1-m)$,
with $\nmax = 10$. }
\label{fig:mu_mn_prolate_ext}
\end{figure}

In the limit $a\to 0$, prolate spheroids get thinner and thinner,
approaching a needle of length $2b$.  In this limit, one has
$\alpha_0\to 0$, and the coefficients $c_{mn}$ diverge in the
leading order as
\begin{equation}
c_{mn} \approx \begin{cases}\frac{1}{a \ln (b/a)} \quad (m = 0),\cr
\frac{|m|}{a} \hskip 10mm (m \ne 0). \end{cases}
\end{equation}
In turn, the coefficients $F_{n,n'}^m(1)$ are finite.  As a
consequence, the eigenvalues $\mu_{mn}$ of the matrix $\MM_m$ diverge
as $a\to 0$ as
\begin{equation}  \label{eq:mu_prolate_a0}
\mu_{mn} \approx \begin{cases} \frac{q_n}{a\, \ln(b/a)} \bigl(1 + O(1/\ln(b/a))\bigr) \quad (m = 0), \cr
\frac{|m|}{a} \hskip 40mm (m\ne 0) , \end{cases}  
\end{equation}
with some prefactors $q_n$.  Figure \ref{fig:mu_mn_prolate_ext}
illustrates the behavior of the first eigenvalues $\mu_{mn}$ as
functions of the minor semi-axis $a$ (with fixed $b = 1$).  At $a =
1$, one retrieves the eigenvalues $(n+1)/b$ for the exterior of a
sphere of radius $b$.  In turn, the eigenvalues $\mu_{mn}$ diverge as
$a\to 0$ according to Eq. (\ref{eq:mu_prolate_a0}).  Note that the
asymptotic behavior (\ref{eq:mu_prolate_a0}) with the numerical
prefactor $q_0 \approx 1$ is quite accurate for $\mu_{00}$.  In turn,
logarithmic corrections to the leading order are more significant for
$\mu_{0n}$ with $n > 0$.

\subsection*{Reflection symmetry}

Since the domain $\Omega$ is symmetric under reflection with respect
to the horizontal plane $z = 0$, the Steklov eigenfunctions inherit
this symmetry.  In fact, if $V(x,y,z)$ is a Steklov eigenfunction,
corresponding to an eigenvalue $\mu$, then $V(x,y,-z)$ is also an
eigenfunction corresponding to the same $\mu$.  Moreover, their linear
combinations $V_\pm(x,y,z) = V(x,y,z) \pm V(x,y,-z)$ are also
eigenfunctions, if they are not zero.  There are thus three options:
(i) $V(x,y,z)$ is symmetric: $V(x,y,z) = V(x,y,-z)$; (ii) $V(x,y,z)$
is antisymmetric: $V(x,y,z) = -V(x,y,-z)$; or (iii) $V(x,y,z)$ is
neither symmetric, nor antisymmetric, in which case $V_+(x,y,z)$ and
$V_-(x,y,z)$ are symmetric and antisymmetric, respectively.

The structure of the matrices $\MM_m$ implies that the Steklov
eigenfunctions $V_{mn}(\alpha,\theta,\phi)$, determined by
Eq. (\ref{eq:Vmn_prolate}), are symmetric (resp., antisymmetric) under
reflection with respect to the horizontal plane $z = 0$ when $m+n$ is
even (resp., odd).  Indeed, the relation (\ref{eq:ALegendre_symm})
causes that the matrix elements $\MM_{mn,mn'}$ are zero when $n+n'$ is
odd.  As a consequence, the elements $[\VV_{mn}]_{mn'}$ of its
eigenvector $\VV_{mn}$ are zero when $n+n'$ is odd, so that
\begin{equation*}
v_{mn}(\pi-\theta,\phi) = \sum\limits_{n'=0}^\infty [\VV_{mn}]_{mn'} (-1)^{m+n'} Y_{mn'}(\theta,\phi),
\end{equation*}
where we used $Y_{mn'}(\pi-\theta,\phi) = (-1)^{m+n'}
Y_{mn'}(\theta,\phi)$ according to Eq. (\ref{eq:ALegendre_symm}).  As
the terms with odd $n+n'$ vanish, one concludes that
\begin{equation}  \label{eq:vmn_symmetry}
v_{mn}(\alpha,\pi-\theta,\phi) = (-1)^{m+n} v_{mn}(\theta,\phi).
\end{equation}
This reflection symmetry is clearly illustrated in
Fig. \ref{fig:vn_prolateExt}.  The same symmetry is preserved for the
Steklov eigenfunctions:
\begin{equation}  \label{eq:Vmn_symmetry}
V_{mn}(\alpha,\pi-\theta,\phi) = (-1)^{m+n} V_{mn}(\alpha,\theta,\phi).
\end{equation}

\subsection*{Half-spheroid in the half-space}

The above symmetries provide a complementary insight onto the Steklov
problem in the upper half-space.  Let
\begin{align*}
\Omega_+ 
& = \biggl\{(x,y,z)\in \R^3 ~:~ \frac{x^2}{a^2} + \frac{y^2}{a^2} + \frac{z^2}{b^2} > 1,~ z > 0 \biggr\} \\
& = \bigl\{ (\alpha,\theta,\phi) ~:~ \alpha > \alpha_0,~ 0 \leq \theta < \pi/2, ~ 0 \leq \phi < 2\pi \bigr\}
\end{align*}
be the exterior of a prolate spheroid in the upper half-space (the
second line highlights that the upper half-space corresponds to $0
\leq \theta < \pi/2$ in the prolate spheroidal coordinates).  The
boundary of this domain is the union of the upper half-spheroidal
surface,
\begin{equation*}
\pa_+ = \bigl\{ (\alpha,\theta,\phi) ~:~ \alpha = \alpha_0,~ 0\leq \theta < \pi/2, ~ 0 \leq \phi < 2\pi \bigr\},
\end{equation*}
and the remaining horizontal plane at $z = 0$ with a circular hole of
radius $a$:
\begin{equation*}
\pa_0 = \bigl\{ (\alpha,\theta,\phi) ~:~ \alpha > \alpha_0,~ \theta = \pi/2, ~ 0 \leq \phi < 2\pi \bigr\}.
\end{equation*}

According to Eq. (\ref{eq:Vmn_symmetry}), antisymmetric Steklov
eigenfunctions $V_{mn}$ (with odd $m+n$) vanish at $\theta = \pi/2$
that corresponds to the Dirichlet boundary condition on $\pa_0$.  As a
consequence, they solve the mixed Steklov-Dirichlet exterior problem:
\begin{subequations}  \label{eq:SteklovD}
\begin{align}
\Delta V_{mn} & = 0  \quad  \textrm{in}~ \Omega_+ , \\ 
\partial_n V_{mn} & = \mu_{mn} V_{mn} \quad \textrm{on} ~ \pa_+, \\
V_{mn} & = 0 \quad \textrm{on}~ \pa_0
\end{align}
\end{subequations}
(here the notation $\partial_n$ for the normal derivative should not
be confused with the index $n$).  Equivalently, one can speak of
eigenvalues $\mu_{mn}$ and eigenfunctions $v_{mn}$ of the
Dirichlet-to-Neumann operator $\M^D$ that maps a given function $f$ on
the half-spheroidal boundary $\pa_+$ onto another function $g = \M^D f
= (\partial_n u)_{|\pa_+}$, where $u$ satisfies
\begin{equation}   \label{eq:Laplace_upper_D}
\Delta u = 0 \quad  \textrm{in}~ \Omega_+,  \qquad u|_{\pa_+} = f, \qquad u|_{\pa_0} = 0.
\end{equation} 
For instance, $v_{01}$, $v_{03}$, $v_{12}$ and $v_{23}$ shown in
Fig. \ref{fig:vn_prolateExt} are examples of eigenfunctions of $\M^D$.

In turn, the symmetric Steklov eigenfunctions $V_{mn}$ (with even
$m+n$) satisfy
\begin{equation*}
\left. \bigl(\partial_n V_{mn}\bigr) \right|_{\pa_0} 
= \left. \biggl(\frac{1}{h_\theta} \, \partial_\theta V_{mn}\biggr)\right|_{\theta =\pi/2} = 0 
\end{equation*}
that corresponds to the Neumann boundary condition on $\pa_0$.  In
other words, they solve the mixed Steklov-Neumann exterior problem:
\begin{subequations}  \label{eq:SteklovN}
\begin{align}
\Delta V_{mn} & = 0  \quad  \textrm{in}~\Omega_+ , \\ 
\partial_n V_{mn} & = \mu_{mn} V_{mn} \quad \textrm{on}~ \pa_+, \\
\partial_n V_{mn} & = 0 \quad \textrm{on}~ \pa_0.
\end{align}
\end{subequations}
Equivalently, one can speak of eigenvalues $\mu_{mn}$ and
eigenfunctions $v_{mn}$ of the Dirichlet-to-Neumann operator $\M^N$
that maps a given function $f$ on the half-spheroidal boundary $\pa_+$
onto another function $g = \M^N f = (\partial_n u)_{|\pa_+}$, where
$u$ satisfies
\begin{equation}      \label{eq:Laplace_upper_N}
\Delta u = 0 \quad  \textrm{in}~ \Omega_+,  \qquad u|_{\pa_+} = f, \qquad (\partial_n u)|_{\pa_0} = 0.
\end{equation} 
For instance, $v_{00}$, $v_{02}$, $v_{11}$, $v_{22}$, $v_{13}$ and
$v_{33}$ shown in Fig. \ref{fig:vn_prolateExt} are examples of
eigenfunctions of $\M^N$.

\section{Oblate spheroids}
\label{sec:oblate}

For the exterior of an oblate spheroid with semi-axes $a \leq b$,
\begin{equation}
\Omega = \biggl\{ (x,y,z)\in\R^3 ~:~ \frac{x^2}{b^2} + \frac{y^2}{b^2} + \frac{z^2}{a^2} > 1 \biggr\} ,
\end{equation}
the computation is very similar so that we only sketch the main steps
and formulas.  In the oblate spheroidal coordinates
$(\alpha,\theta,\phi)$,
\begin{equation*}
\left(\begin{array}{c} x \\ y \\ z \\ \end{array}\right) = a_E \left(\begin{array}{c} 
\cosh \alpha \cos \theta \cos\phi \\ 
\cosh \alpha \cos \theta \sin\phi \\ 
\sinh \alpha \sin \theta  \\  \end{array} \right)   \quad
\left\{\begin{array}{l} 0 \leq \alpha < \infty \\ -\tfrac{\pi}{2} \leq \theta \leq \tfrac{\pi}{2} \\ 0 \leq \phi < 2\pi \\ \end{array} \right\},
\end{equation*}
with $a_E = \sqrt{b^2 - a^2}$ (Fig. \ref{fig:scheme}b), the scale
factors determining the surface and volume elements, are
\cite{Korn}
\begin{align*}
h_\alpha & = h_\theta = a_E \sqrt{\sinh^2\alpha + \sin^2 \theta}, \\
h_\phi & = a_E \cosh\alpha \cos\theta .
\end{align*}
In these coordinates, the domain $\Omega$ is still characterized by
$\alpha > \alpha_0 = \tanh^{-1}(a/b)$, and the action of the
Laplace operator reads
\begin{align} \nonumber
\Delta u & = \frac{1}{a_E^2 (\sinh^2\alpha + \sin^2\theta)} \biggl[\frac{1}{\cosh\alpha} 
\frac{\partial}{\partial \alpha} \biggl(\cosh\alpha \frac{\partial u}{\partial \alpha}\biggr) \\  \label{eq:Laplace_pblate}
& + \frac{1}{\cos\theta} \frac{\partial}{\partial \theta} \biggl(\cos\theta \frac{\partial u}{\partial \theta}\biggr)
\biggr] + \frac{1}{a_E^2 \cosh^2\alpha \, \cos^2\theta} \frac{\partial^2 u}{\partial \phi^2} \,.
\end{align}
A general solution of the Laplace equation reads
\begin{equation}  \label{eq:u_oblate}
u(\alpha,\theta,\phi) = \sum\limits_{n=0}^\infty \sum\limits_{m=-n}^n A_{mn} Q_n^m(i\sinh \alpha)
\bar{Y}_{mn}(\theta,\phi) ,
\end{equation}
where $\bar{Y}_{mn}(\theta,\phi) = a_{mn} P_n^m(\sin\theta)
e^{im\phi}$, and the action of the Dirichlet-to-Neumann operator is
\begin{align}  \label{eq:Mu_oblate}
& \M u|_{\pa} = \left. \frac{\partial u}{\partial n} \right|_{\pa} 
= - \left. \frac{1}{h_\alpha} \partial_\alpha u  \right|_{\alpha=\alpha_0}  \\  \nonumber
& = \sum\limits_{n=0}^\infty \sum\limits_{m=-n}^n A_{mn}  \bar{Y}_{mn}(\theta,\phi) 
\frac{\cosh \alpha_0 \, Q^{'m}_n(i\sinh \alpha_0)}{i a_E \sqrt{\cosh^2\alpha_0 - \cos^2\theta}} \,.
\end{align}
Multiplying this relation by $\bar{Y}_{m'n'}^*(\theta,\phi) \cos
\theta$ and integrating over $\theta$ and $\phi$, one gets a matrix
representation of the operator $\M$ on the orthonormal basis of
$\bar{Y}_{mn}$:
\begin{equation}  \label{eq:M_oblate}
\MM_{mn,m'n'} = 2\pi\delta_{m,m'} a_{mn} a_{m'n'} \, c_{m'n'} \, \bar{F}_{n,n'}^m(\sinh\alpha_0),
\end{equation}
where
\begin{equation}  \label{eq:cmn_oblate}
c_{mn} = \frac{\cosh \alpha_0 \, Q^{'m}_n(i\sinh \alpha_0)}{i a_E Q_n^m(i\sinh\alpha_0)} 
\end{equation}
and
\begin{equation} \label{eq:Fmn_oblate}
\bar{F}_{n,n'}(z) = \int\limits_{-1}^1 dx \frac{P_n^{m}(x) P_{n'}^m(x)}{\sqrt{z^2 + x^2} } .
\end{equation} 
According to Eq. (\ref{eq:Fmnn}), one also has
\begin{equation} \label{eq:barFmn_Fmn}
\bar{F}_{n,n'}^m(z) = i F_{n,n'}^m(iz).
\end{equation}

Since the structure of the matrix $\MM$ is the same as for prolate
spheroids, many properties of eigenfunctions of the
Dirichlet-to-Neumann operator remain unchanged; in particular, they
can be classified according to their dependence on $\phi$ via
$e^{im\phi}$; we employ the double index $mn$ in the following.  Using
again the decomposition (\ref{eq:M_sum}), one can diagonalize
separately the matrices $\MM_m$ to access the eigenvalues $\mu_{mn}$
and to construct the eigenfunctions
\begin{equation}  \label{eq:vmn_oblate}
v_{mn}(\theta,\phi) = \sum\limits_{n'=0}^\infty [\VV_{mn}]_{mn'} \bar{Y}_{mn'}(\theta,\phi)
\end{equation} 
and the Steklov eigenfunctions
\begin{equation}
V_{mn}(\alpha,\theta,\phi) = \sum\limits_{n'=0}^\infty \frac{Q_{n'}^m(i\sinh \alpha)}{Q_{n'}^m(i\sinh \alpha_0)} 
[\VV_{mn}]_{mn'}  \bar{Y}_{mn'}(\theta,\phi).
\end{equation} 
As previously, the Steklov eigenfunctions $V_{mn}$ are symmetric for
even $m+n$ and antisymmetric for odd $m+n$ under reflection with
respect to the horizontal plane $z = 0$.  In particular, these
eigenfunctions solve the mixed Steklov-Neumann and Steklov-Dirichlet
exterior problems for the exterior of an oblate spheroid in the upper
half-space.

\begin{figure}
\begin{center}
\includegraphics[width=88mm]{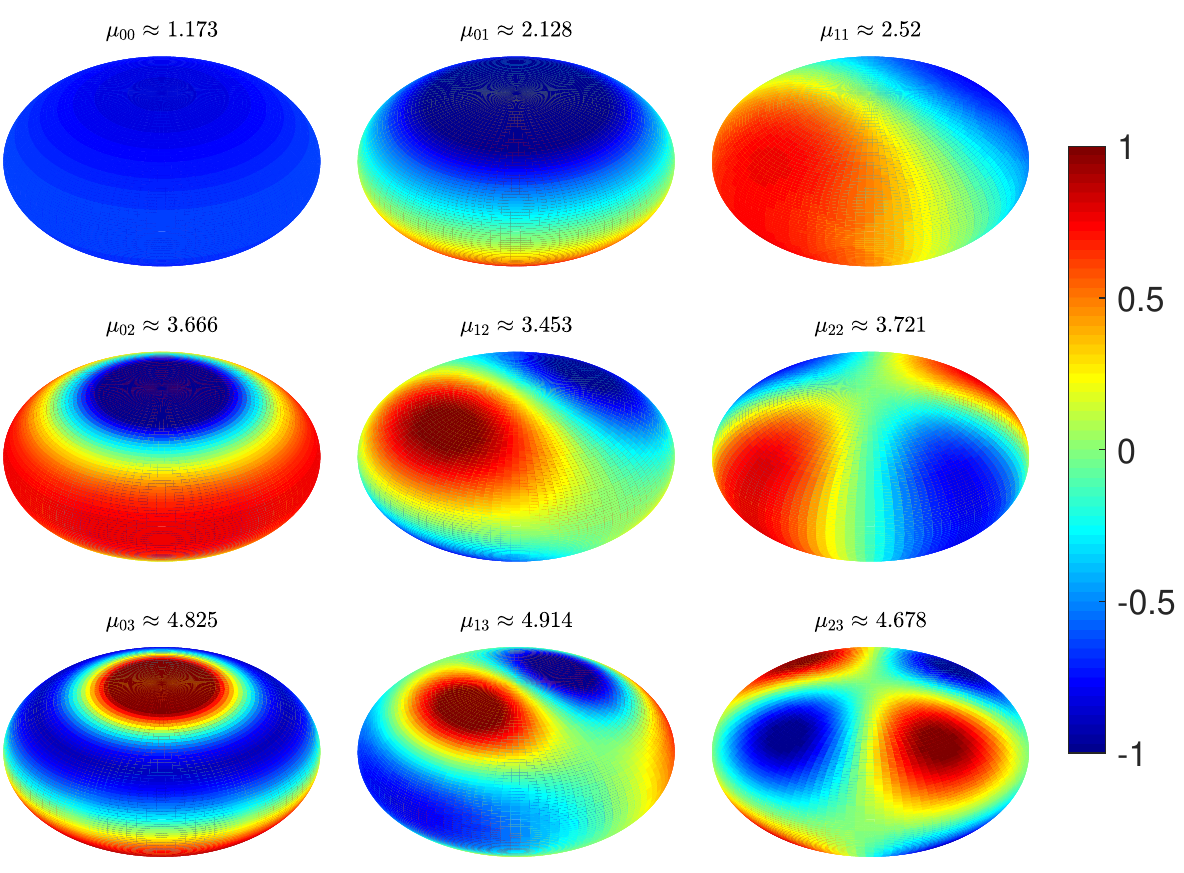} 
\end{center}
\caption{
Several eigenfunctions $v_{mn}$ of the Dirichlet-to-Neumann operator
$\M$ for the exterior of the oblate spheroid with semi-axes $a = 0.5$
and $b = 1$.  The associated eigenvalues are shown on the top.  The
eigenfunctions with even $m+n$ are symmetric (with respect to the
horizontal plane $z = 0$), whereas the eigenfunctions with odd $m+n$
are antisymmetric, in agreement with Eq. (\ref{eq:vmn_symmetry}).  The
truncation order is $\nmax = 10$.}
\label{fig:vn_oblateExt}
\end{figure}

Figure \ref{fig:vn_oblateExt} illustrates several eigenfunctions
$v_{mn}$ of the Dirichlet-to-Neumann operator $\M$ for the exterior of
the oblate spheroid with semi-axes $a = 0.5$ and $b = 1$.  Expectedly,
these eigenfunctions resemble spherical harmonics $\bar{Y}_{mn}$ but
exhibit some differences.  In particular, the ground eigenfunction
$v_{00}$ is not constant, though its variations are small.

\begin{figure}
\begin{center}
\includegraphics[width=88mm]{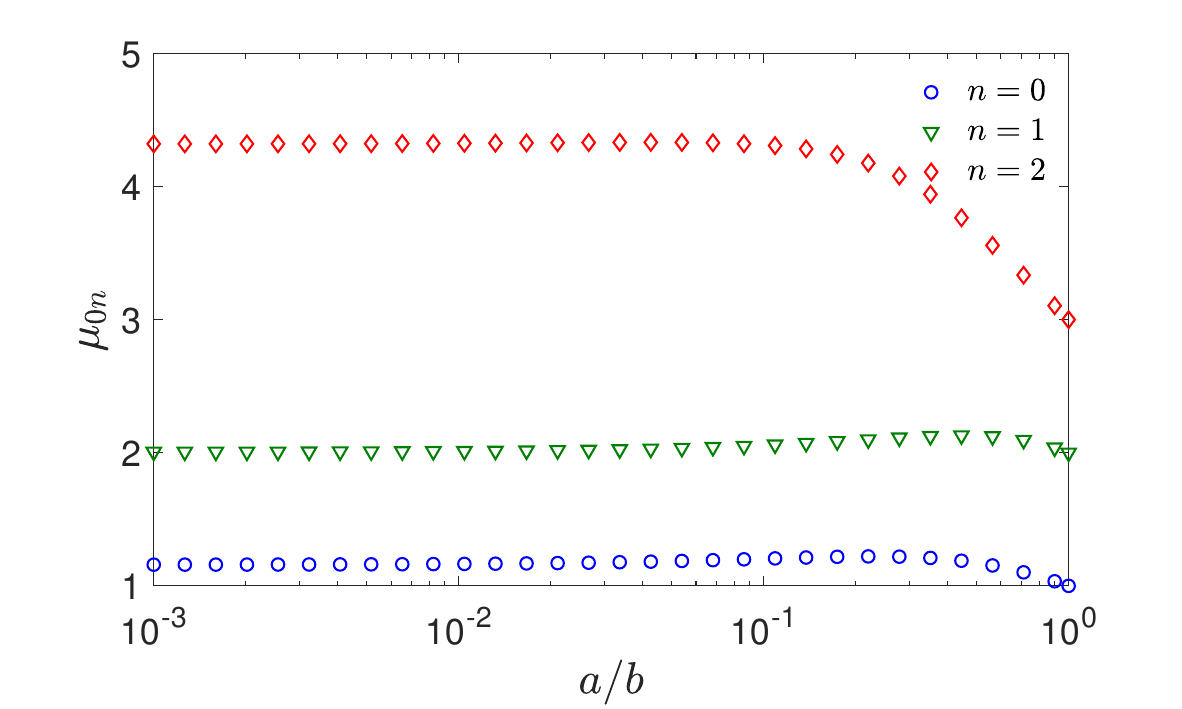} 
\end{center}
\caption{
Three eigenvalues $\mu_{0n}$ of the Dirichlet-to-Neumann operator for
the exterior of an oblate spheroid as functions of its minor semi-axis
$a$, with $b = 1$.  At $a = b$, one retrieves the eigenvalues
$(n+1)/b$ for the exterior of a sphere of radius $b$.  As $a\to 0$,
one gets the eigenvalues for the exterior of a disk of radius $b$.
The axisymmetric matrix $\MM_0$ from Eq. (\ref{eq:M_oblateA}) was
truncated to the size $(\nmax+1)\times (\nmax+1)$, with $\nmax = 20$.
}
\label{fig:muk_oblate_ext}
\end{figure}

Figure \ref{fig:muk_oblate_ext} illustrates the behavior of three
eigenvalues $\mu_{0n}$ as functions of the minor semi-axis $a$ (with
fixed $b = 1$).  At $a = 1$, one retrieves the eigenvalues $(n+1)/b$
for the exterior of a sphere of radius $b$.  As $a$ decreases, oblate
spheroids become thinner and thinner, approaching a disk of radius
$b$.  In contrast to the case of prolate spheroids, the eigenvalues
are finite in this limit.  Curiously, each eigenvalue does not change
monotonously with $a$.  Table \ref{tab:mu_oblateExt} summarizes
several eigenvalues $\mu_{mn}$ for oblate spheroids with the minor
semi-axis $a$ ranging from $0$ to $1$ (with $b = 1$).

\begin{table}
\begin{tabular}{|c|c|c|c|c|c|c|}  \hline
$a/b$ & $\mu_{00}$ & $\mu_{01}$ & $\mu_{02}$ & $\mu_{11}$ & $\mu_{12}$ & $\mu_{22}$ \\ \hline
0.0  &  1.158  &  2.006 &  4.317 &  2.755 &  3.453 &  4.121  \\
0.1  &  1.204  &  2.057 &  4.314 &  2.811 &  3.512 &  4.197 \\
0.2  &  1.220  &  2.094 &  4.206 &  2.796 &  3.539 &  4.166 \\
0.3  &  1.217  &  2.117 &  4.040 &  2.732 &  3.536 &  4.058 \\
0.4  &  1.200  &  2.129 &  3.850 &  2.634 &  3.506 &  3.900 \\
0.5  &  1.173  &  2.128 &  3.666 &  2.520 &  3.453 &  3.721 \\
0.6  &  1.141  &  2.118 &  3.498 &  2.401 &  3.381 &  3.543 \\
0.7  &  1.106  &  2.098 &  3.350 &  2.286 &  3.295 &  3.379 \\
0.8  &  1.070  &  2.071 &  3.220 &  2.180 &  3.200 &  3.234 \\
0.9  &  1.034  &  2.038 &  3.105 &  2.085 &  3.101 &  3.108 \\
1.0  &  1      &  2     &  3     &  2     &  3    &   3     \\ \hline
\end{tabular}
\caption{
Several eigenvalues $\mu_{mn}$ of the Dirichlet-to-Neumann operator
$\M$ for the exterior of the oblate spheroid with semi-axes $a$ and $b
= 1$.  The matrices $\hat{\MM}_m$ are truncated to the size $(\nmax+1
- m)\times (\nmax+1 - m)$, with the truncation order $\nmax = 10$.
Further increase of $\nmax$ or decrease of $\delta$ do not change the
shown eigenvalues.  }
\label{tab:mu_oblateExt}
%
\end{table}

\subsection*{Limit of a disk}

Let us focus on the limit of the disk when $a = 0$ and thus $\alpha_0
= 0$.  One can introduce the radial coordinate is $r = \sqrt{x^2+y^2}
= b |\cos\theta|$, with $0 < \theta < \pi/2$ corresponding to the
upper side of the disk, and $-\pi/2 < \theta < 0$ corresponding to its
lower side.  Formally, one can set $r = b \cos\theta$ and associate
positive $r$ with the upper side of the disk and negative $r$ with its
lower side.

As $\alpha_0 \to 0$, the functions $\bar{F}_{n,n'}^m(z)$ from
Eq. (\ref{eq:Fmn_oblate}) logarithmically diverge in the limit $z =
\sinh\alpha_0 \to 0$.  For any $\alpha_0 > 0$, one can still use
the truncated matrix $\MM$ (or $\MM_m$) to approximate the eigenvalues
and eigenfunctions but larger and larger truncation orders are needed
as $\alpha_0$ decreases.  For this reason, it is convenient to use an
alternative matrix representation, which corresponds to the operator
$\M^{-1}$ and remains valid even in the limit $\alpha_0 = 0$.  This
representation is described in Appendix \ref{sec:alternative}.

Figure \ref{fig:vn_diskExt} illustrates several eigenfunctions
$v_{mn}$ of the Dirichlet-to-Neumann operator for the exterior of the
disk of radius $b = 1$.  One can see that the ground eigenfunction
$v_{00}$ shows more significant variations as compared to that shown
in Fig. \ref{fig:vn_oblateExt}.  We stress that the eigenfunctions are
also present on the bottom side of the disk, which is not visible.
Their structure on this hidden side can be easily reconstructed from
their symmetry with respect to the horizontal plane: $v_{mn}$ is
symmetric (resp., antisymmetric) for even (resp., odd) $m+n$.  For
instance, the eigenfunction $v_{01}$, which is positive on the upper
side of the disk, takes negative values on the bottom side.

\begin{figure}
\begin{center}
\includegraphics[width=88mm]{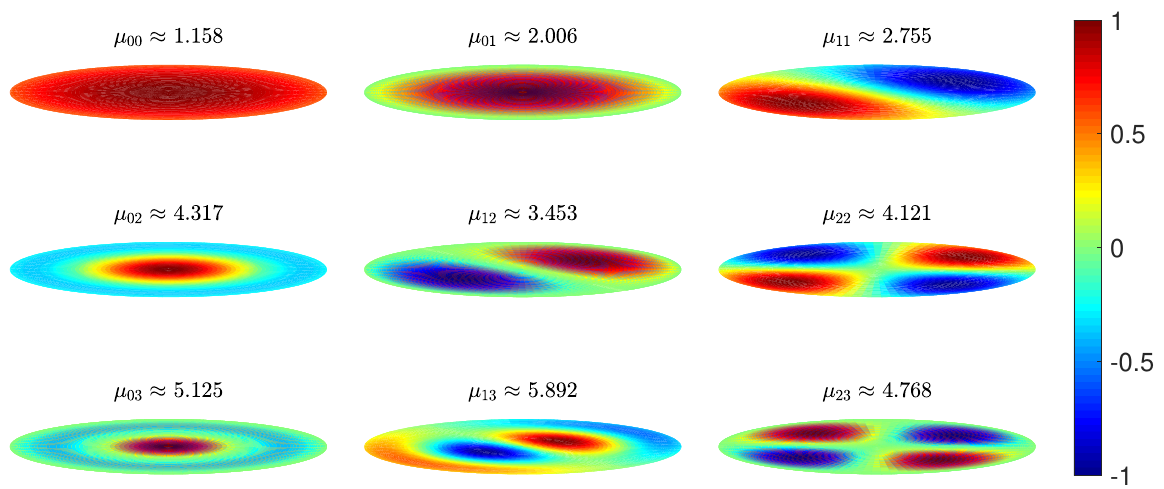} 
\end{center}
\caption{
Several eigenfunctions $v_{mn}$ of the Dirichlet-to-Neumann operator
$\M$ for the exterior of the disk of radius $b = 1$.  The associated
eigenvalues are shown on the top.  Normalization by the maximum of
$|v_{mn}|$ was employed for a better visualization.  The symmetric
(resp., antisymmetric) shape of the eigenfunction $v_{mn}$ for even
(resp., odd) $m+n$ allows one to reconstruct its structure on the
bottom side of the disk, which is not visible.  The truncation order
is $\nmax = 10$. }
\label{fig:vn_diskExt}
\end{figure}

\subsubsection*{Axisymmetric setting}

Let us briefly discuss the action of the Dirichlet-to-Neumann operator
onto rotationally invariant functions that do not depend on the angle
$\phi$.  If a function $f(\theta)$ is decomposed on the normalized
Legendre polynomials $\psi_n(\theta) = \sqrt{n+1/2} \,
P_n(\sin\theta)$,
\begin{equation}
f(\theta) = \sum\limits_{n=0}^\infty f_n \psi_n(\theta) ,
\end{equation}
the action of $\M$ reads
\begin{equation}
\M f = \sum\limits_{n=0}^\infty f_n \psi_n(\theta) \frac{c_{0n}}{b \sqrt{1-\cos^2\theta}} \,.
\end{equation}
Setting $b = 1$ for simplicity and using $r = \cos\theta$, we get, for
instance,
\begin{subequations}
\begin{align}  \label{eq:Weber}
\M 1 & = \frac{c_{00}}{\sqrt{1-r^2}} \,,\\
\M \sqrt{1-r^2} & = c_{01} ,
\end{align}
\end{subequations}
where 
\begin{equation}   \label{eq:c0n_disk}
c_{0n} = \frac{-i Q'_n(0)}{Q_n(0)} = 2\biggl(\frac{\Gamma(n/2+1)}{\Gamma(n/2+1/2)}\biggr)^2 
\end{equation}
(see Appendix \ref{sec:axi} for details).  For instance, $c_{00} =
2/\pi$, $c_{01} = \pi/2$, $c_{02} = 8/\pi$, $c_{03} = 9\pi/8$, etc.
The first relation (\ref{eq:Weber}) reproduces the classical Weber's
solution for the electric current density onto a conducting disk
\cite{Sneddon}, see below.

Alternatively, setting $x = \sin\theta$, one has
\begin{equation}
\M P_n(x) = c_{0n} \frac{P_n(x)}{|x|} \qquad (-1 < x < 1).
\end{equation}
One sees that $\M P_n(x)$ keeps the parity of $P_n(x)$, i.e., it is
symmetric for even $n$ and antisymmetric for odd $n$.  For instance,
one has
\begin{align*}
\M 1 & = \frac{2}{\pi}\, \frac{1}{|x|} \,, \\
\M x & = \frac{\pi}{2}\, \frac{x}{|x|} \,, \\
\M x^2 & = \frac{2}{\pi} \, \frac{4x^2 - 1}{|x|} \,. 
\end{align*}

\begin{figure}
\begin{center}
\includegraphics[width=88mm]{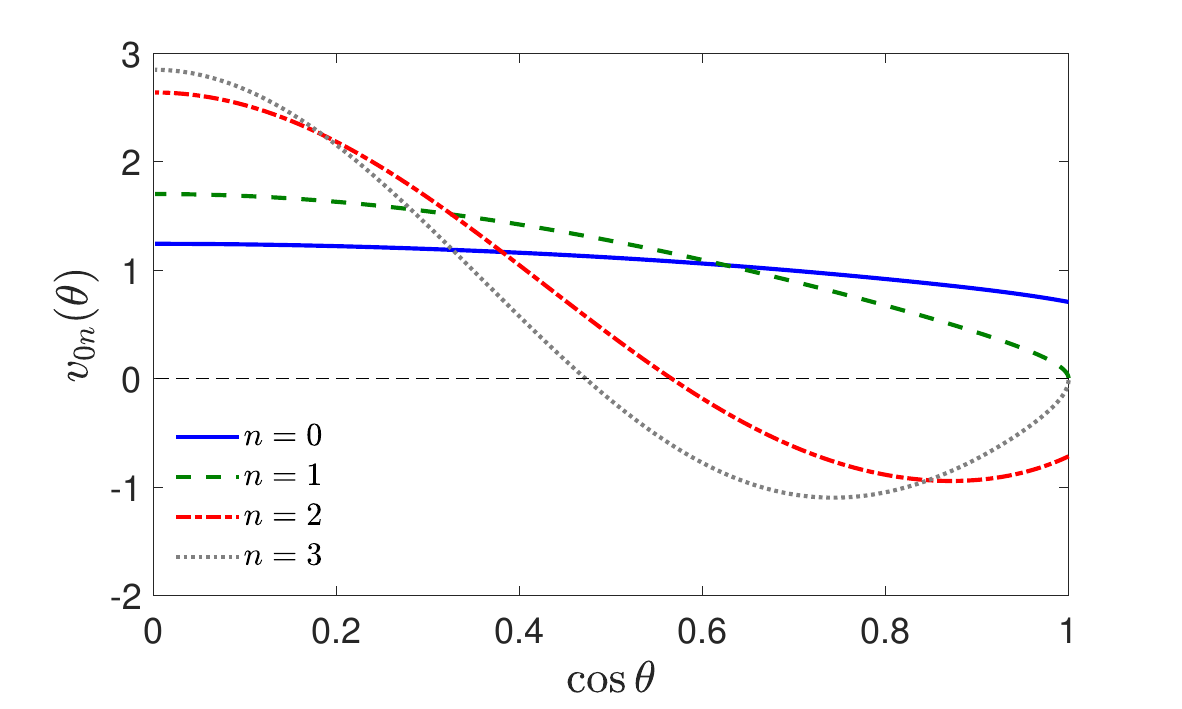} 
\end{center}
\caption{
Four eigenfunctions $v_{0n}(\theta)$ of the Dirichlet-to-Neumann
operator for the exterior of a disk of radius $b = 1$, as a function
of $r = b \cos\theta$.  The associated eigenvalues are $1.158$,
$2.006$, $4.317$, and $5.125$.  The axisymmetric matrix $\MM_0$ is
truncated to $(\nmax+1) \times (\nmax+1)$, with $\nmax = 20$.  }
\label{fig:vk_disk}
\end{figure}

\subsubsection*{Disk in the half-space}

The general representation (\ref{eq:u_oblate}) allows one to solve
boundary value problems (\ref{eq:Laplace_upper_D},
\ref{eq:Laplace_upper_N}) in the upper half-space, by imposing
Dirichlet or Neumann boundary condition on $\pa_0$.  In the case of a
disk of radius $b$, the first problem actually concerns the Dirichlet
boundary condition on the horizontal plane, for a class of functions
$f(r)$ that are strictly zero for $r \geq b$.  Its solution can be
written in the integral form:
\begin{align} \nonumber
u(x_0,y_0,z_0) & = \int\limits_{x^2+y^2 < b^2} dx \, dy \, f(x,y) \\
& \times \frac{z_0}{2\pi(z_0^2 + (x-x_0)^2 + (y-y_0)^2)^{3/2}} \,,
\end{align}
where the factor in front of $f$ is the harmonic measure density
\cite{Garnett}.  Alternatively, one can use the representation
(\ref{eq:u_oblate}) of a harmonic function in oblate spheroidal
coordinates, in which the coefficients $A_{mn}$ are obtained by
setting $\alpha = 0$, multiplying by $\bar{Y}_{mn}^* \cos\theta$ and
integrating over $\theta$ and $\phi$:
\begin{equation}  \label{eq:Amn_disk}
A_{mn} = \frac{1}{Q_n^m(0)} \int\limits_{-\pi/2}^{\pi/2} d\theta \int\limits_0^{2\pi} d\phi \,
f(\theta,\phi) \, \bar{Y}_{mn}^*(\theta,\phi) \cos\theta.
\end{equation}
For instance, if  $f(\theta,\phi) = \textrm{sign}(\theta)$, one gets
\begin{align}  \nonumber
A_{mn} &= \frac{2\pi \delta_{m,0} a_{0n}}{Q_n^m(0)} \int\limits_{-1}^{1} dx\,  \textrm{sign}(x) \, P_n(x) \\
& = \frac{2\pi \delta_{m,0} a_{0n}}{Q_n^m(0)} \, \frac{P_{n-1}(0)-P_{n+1}(0)}{n+1/2} 
\end{align}
for odd $n$, where $P_{2n}(0) = \frac{(-1)^n (2n)!}{2^{2n} (n!)^2}$.
As a consequence, one has
\begin{equation}
u = \sum\limits_{n=0}^\infty \frac{Q_{2n+1}(i\sinh\alpha)}{Q_{2n+1}(0)}  P_{2n+1}(\sin\theta)
\bigl(P_{2n}(0)-P_{2n+2}(0)\bigr) \,.
\end{equation}
This is also the solution of the Laplace equation in the upper
half-space with the boundary condition $u|_{z=0} = \Theta(b-r)$, where
$\Theta(z)$ is the Heaviside step function: $\Theta(z) = 1$ for $z >
0$ and $0$ otherwise.  This is the splitting probability for the disk
of radius $b$, i.e., the probability that a particle started from a
point $(\alpha,\theta,\phi)$ hits the disk before hitting its
complement (the horizontal plane without this disk)
\cite{Garnett,VanKampen,Redner}.

In the same vein, the representation (\ref{eq:u_oblate}) with even
$m+n$ allows one to solve the mixed Dirichlet-Neumann boundary value
problem (\ref{eq:Laplace_upper_N}) in the upper half-space, with
coefficients $A_{mn}$ given by Eq. (\ref{eq:Amn_disk}).  For instance,
if $f = 1$, one gets
\begin{equation}
A_{mn} = \frac{2\sqrt{\pi} \delta_{m,0} \delta_{n,0}}{Q_n(0)} \,,
\end{equation}
so that
\begin{equation}   \label{eq:u_f1}
u = \frac{Q_0(i\sinh\alpha)}{Q_0(0)} \,,
\end{equation}
where $\alpha = \cosh^{-1}((r_+ + r_-)/(2b))$, with $r_{\pm} =
\sqrt{(r\pm b)^2 + z^2}$.  This expression can be written more
explicitly as
\begin{equation}  \label{eq:u_f1bis}
u(r,z) = \frac{1}{\pi} \cos^{-1} \biggl(1 - \frac{8b^2}{(r_+ + r_-)^2}\biggr).
\end{equation}
It is worth noting that the mixed Dirichlet-Neumann boundary value
problem for rotationally invariant functions $f(r)$,
\begin{equation}
\Delta u = 0 \quad (z > 0), \quad \left\{ \begin{array}{ll} u(r,0) = f(r) & (0 \leq r < b), \\
(\partial_z u(r,z))|_{z=0} = 0 & (r > b) , \\ \end{array} \right. 
\end{equation}
was solved by Beltrami via integral representations (see
\cite{Sneddon}, page 66):
\begin{equation}
u(r,z) = \int\limits_0^\infty d\xi\, A(\xi)\, e^{-z\xi}\, J_0(\xi r) ,
\end{equation}
where
\begin{equation}
A(\xi) = \int\limits_0^1 ds  \, \cos(\xi s)  \frac{d}{ds} 
\left( \frac{2}{\pi} \int\limits_0^s dt \, \frac{t \, f(tb)}{\sqrt{s^2-t^2}} \right).
\end{equation}
For instance, setting $f(r) = 1$, one retrieves the Weber's solution
for the electric potential around a conducting disk:
\begin{equation}   \label{eq:u_f1int}
u(r,z) = \frac{2}{\pi}\int\limits_0^\infty \frac{d\xi}{\xi} \sin(\xi b) \, e^{-\xi z} \, J_0(\xi r),
\end{equation}
from which
\begin{equation}
-(\partial_n u)|_{\pa_+} = \frac{2}{\pi \sqrt{b^2 - r^2}}  \qquad  (0 \leq r < b),
\end{equation}
where $J_0(z)$ is the Bessel function of the first kind.  One sees
that the above expressions (\ref{eq:u_f1}) and (\ref{eq:u_f1bis}) are
more explicit than the equivalent integral representation
(\ref{eq:u_f1int}).  In probabilistic terms, $1-u(r,z)$ is the escape
(or survival) probability for a particle started from a point $(r,z)$,
which can be destroyed upon its first arrival onto the absorbing disk.
Alternatively, $C_0 u(r,z)$ can be understood as the concentration of
particles diffusing from infinity (with a concentration $C_0$) towards
an absorbing disk.  In this light, $j = - D C_0 (\partial_n
u)|_{\pa_+}$ is the diffusive flux density onto the disk.  Dividing
this expression by the total flux, one gets the harmonic measure
density on the absorbing disk for a particle started from infinity:
\begin{equation}
\omega(r) = \frac{DC_0 \frac{2}{\pi \sqrt{b^2 - r^2}}}{D C_0 \frac{2}{\pi} b} = \frac{1}{b \sqrt{b^2 - r^2}} \,.
\end{equation}

\section{Two applications}
\label{sec:application}

As mentioned in Sec. \ref{sec:intro}, the Dirichlet-to-Neumann
operator $\M$ and its eigenfunctions have found numerous applications
in applied mathematics and engineering.  Here we briefly discuss two
applications in physics.  In Sec. \ref{sec:Robin}, we start with
diffusion-controlled reactions in chemical physics and illustrate the
use of $\M$ for representing the Robin boundary condition that is
often used to describe surface reactions on partially reactive
targets.  In Sec. \ref{sec:encounters}, we highlight their relation to
the statistics of encounters between a diffusing particle and the
boundary, which is relevant in statistical physics and in the theory
of reflected Brownian motion.

\subsection{Partially reactive boundaries}
\label{sec:Robin}

Diffusion-controlled reactions play an important role in physics,
chemistry and biology
\cite{Schuss,Metzler,Lindenberg,North66,Wilemski73,Calef83,Berg85,Rice85,Bressloff13,Benichou14,Grebenkov23n}.
In the idealized setting introduced by Smoluchowski
\cite{Smoluchowski17}, a reactant diffuses towards its partner or
towards a catalytic surface, and reacts upon their first encounter.
In many situations, however, the first encounter is not sufficient, as
the reactant and/or its partner may need to overcome an activation
energy barrier, to be in an appropriate ``active'' state, to arrive
onto a specific catalytic germ, to pass through a channel/pore,
etc. \cite{Grebenkov23n}.  If any of these conditions is failed, the
reactant resumes its bulk diffusion until the next encounter, and so
on.  Starting from Collins and Kimball \cite{Collins49}, such partial
reactions are described by imposing the Robin boundary condition, in
which the diffusive flux of particles towards the catalytic boundary
is {\it postulated} to be proportional to their concentration on that
boundary.  In other words, the concentration $C$ of diffusing
reactants in the steady-state regime obeys:
\begin{subequations}
\begin{align}
\Delta C & = 0 \quad \textrm{in}~\Omega, \\  \label{eq:C_Robin}
-D \partial_n C & = \kappa C  \quad \textrm{on}~ \pa , \\
\lim\limits_{|\x|\to\infty} C & = C_0,
\end{align}
\end{subequations} 
where $D$ is the diffusion coefficient, $\kappa \geq 0$ is a constant
reactivity of the boundary, and $C_0$ is the initial concentration
maintained at infinity.  Setting $q = \kappa/D$ and $C = C_0(1-u)$,
one can express the Robin boundary condition $\partial_n u + q u = q$
on $\pa$ with the help of the Dirichlet-to-Neumann operator as $\M
u|_{\pa} + q u|_{\pa} = q$.  Inverting this relation yields
\begin{align}  \nonumber
u|_{\pa} & = (\M + q)^{-1} q \\
& = \sum\limits_{k=0}^\infty \frac{(v_k,1)_{L^2(\pa)}}{\|v_k\|^2_{L^2(\pa)}} \, \frac{q}{\mu_k+q} \,v_k ,
\end{align}
where we used the orthogonality and completeness of the basis of the
eigenfunctions $\{v_k\}$ of $\M$.

For prolate spheroids, a general solution (\ref{eq:u_prolate}) of the
Laplace equation can be written as
\begin{equation}
u(\alpha,\theta,\phi) = \sum\limits_{m,n} f_{mn} \frac{Q_n^m(\cosh\alpha)}{Q_n^m(\cosh\alpha_0)} Y_{mn}(\theta,\phi) ,
\end{equation}
where the coefficients $f_{mn}$ are found from the restriction of $u$
onto $\pa$ as:
\begin{align} \nonumber
f_{mn} & = \int\limits_0^\pi d\theta \int\limits_0^{2\pi} d\phi  \, \sin\theta\,  Y_{mn}^*(\theta,\phi) \, u|_{\pa} \\ \nonumber
& = \sum\limits_{k=0}^\infty \frac{(v_k,1)_{L^2(\pa)}}{\|v_k\|^2_{L^2(\pa)}} \, \frac{q}{\mu_k+q} \\ \nonumber
& \times \int\limits_0^\pi d\theta \int\limits_0^{2\pi} d\phi  \, \sin\theta \, Y_{mn}^*(\theta,\phi) \, v_k(\theta,\phi) \\
& = \sum\limits_{k=0}^\infty \frac{(v_k,1)_{L^2(\pa)}}{\|v_k\|^2_{L^2(\pa)}} \, \frac{q}{\mu_k+q}  [\VV_k]_{mn},
\end{align}
where we employed the representation (\ref{eq:vk_prolate}).  Using
Eqs. (\ref{eq:vknorm_prolate}, \ref{eq:vk_1_prolateE}) from Appendix
\ref{sec:numerics}, one can further simplify:
\begin{align}  \nonumber
f_{mn} & = \frac{\sqrt{4\pi}}{a_E \sinh\alpha_0 Q_0(\cosh\alpha_0)}
\sum\limits_{k=0}^\infty \frac{[\VV_k^*]_{00} [\VV_k]_{mn}}{(\VV_k^\dagger \cc \VV_k)}  \,\frac{q}{\mu_k+q} \\
& = \sqrt{4\pi} \sum\limits_{k=0}^\infty \frac{[\tilde{\VV}_k^*]_{00} 
[\tilde{\VV}_k]_{mn}}{(\tilde{\VV}_k^\dagger \tilde{\VV}_k)}  \,\frac{q}{\mu_k+q}  \,,
\end{align}
where $\tilde{\VV}_k = \cc^{\frac12} \VV_k$, and the diagonal matrix
$\cc$ is formed by $c_{mn}$.  Since $\mu_k$ and $\tilde{\VV}_k$ are
eigenvalues and eigenvectors of the Hermitian matrix $\tilde{\MM} =
\cc^{\frac12} \MM \cc^{-\frac12}$, one can also rewrite the above
expression in a matrix form:
\begin{equation}  \label{eq:fmn_M}
f_{mn} = \sqrt{4\pi} \, q \bigl[(\tilde{\MM} + q\II)^{-1}\bigr]_{mn,00} ,
\end{equation}
where $\II$ is the identity matrix.  Similar representations were
discussed in \cite{Piazza19,Grebenkov19}.

Knowing the concentration, one can easily deduce the total diffusive
flux onto the boundary:
\begin{align}  \nonumber
J_q & = \int\limits_{\pa} (-D\partial_n C) = D C_0 \int\limits_{\pa} (\partial_n u)|_{\pa} \\  \nonumber
& = D C_0 q \bigl(1, \M (\M + q)^{-1} 1\bigr)_{L^2(\pa)} \\  \nonumber
& = D C_0 q \sum\limits_{k=0}^\infty \frac{|(1, v_k)_{L^2(\pa)}|^2}{\|v_k\|^2_{L^2(\pa)}} \, \frac{\mu_k}{\mu_k + q} \\  \nonumber
& = \frac{4\pi D C_0 q}{\sinh\alpha_0 Q_0^2(\cosh\alpha_0)} 
\sum\limits_{k=0}^\infty \frac{ [\VV_k^*]_{00} \, [\VV_k]_{00}}{(\VV_k^\dagger \cc \VV_k)} \, \frac{1}{\mu_k + q} \\ \nonumber
& = \frac{4\pi D C_0 a_E q}{Q_0(\cosh\alpha_0)} 
\sum\limits_{k=0}^\infty \frac{ [\tilde{\VV}_k^*]_{00} \, [\tilde{\VV}_k]_{00}}{(\tilde{\VV}_k^\dagger \tilde{\VV}_k)} \, \frac{1}{\mu_k + q} \\
\label{eq:Jq_prolate}
& = \frac{4\pi D C_0 a_E q}{Q_0(\cosh\alpha_0)}  \bigl[(\tilde{\MM} + q\II)^{-1}\bigr]_{00,00} .
\end{align}
Denoting by
\begin{equation}  \label{eq:Charm_prolate}
\C = \frac{4\pi a_E}{Q_0(\cosh\alpha_0)} = \frac{8\pi a_E}{\ln \bigl(\frac{b+a_E}{b-a_E}\bigr)}
\end{equation}
the harmonic (Newtonian) capacity of the prolate spheroid, one gets
\begin{equation}
J_q = J_\infty \, q  \bigl[(\tilde{\MM} + q\II)^{-1}\bigr]_{00,00} ,
\end{equation}
where 
\begin{equation}
J_\infty = D C_0 \C \,
\end{equation}
is the total diffusive flux onto a perfectly reactive prolate spheroid
(see also \cite{Piazza19}).
%
As a consequence, the effect of partial reactivity is represented by
$q\bigl[(\tilde{\MM} + q\II)^{-1}\bigr]_{00,00}$.  In the limit of a
sphere ($a\to b$), this factor reduces to $q/(q + 1/b)$ so that one
retrieves the Collins-Kimball diffusive flux \cite{Collins49}
\begin{equation}
J_q = 4\pi D C_0 b  \, \frac{qb}{1 + qb}  \qquad (a=b),
\end{equation}
where $4\pi DC_0 b$ is the Smoluchowski diffusive flux $J_\infty$ onto
a perfectly reactive sphere of radius $b$ \cite{Smoluchowski17}.

For oblate spheroids, a similar computation involves
\begin{equation}
u(\alpha,\theta,\phi) = \sum\limits_{m,n} f_{mn} \frac{Q_n^m(i\sinh\alpha)}{Q_n^m(i\sinh\alpha_0)} \bar{Y}_{mn}(\theta,\phi) ,
\end{equation} 
with the same expression (\ref{eq:fmn_M}) for the coefficients
$f_{mn}$.  The total diffusive flux reads
\begin{equation}
\label{eq:Jq_oblate}
J_q = J_\infty ~ q \bigl[(\tilde{\MM} + q\II)^{-1}\bigr]_{00,00} \,,
\end{equation}
where 
\begin{equation}
J_\infty = \frac{4\pi D C_0 a_E}{iQ_0(i\sinh\alpha_0)} = D C_0 \C, 
\end{equation}
and 
\begin{equation}   \label{eq:Charm_oblate}
\C = \frac{4\pi a_E}{iQ_0(i\sinh\alpha_0)} = \frac{4\pi a_E}{\cos^{-1}(a/b)}
\end{equation}
is the harmonic capacity of the oblate spheroid.  In the case of a
disk, the total flux was thoroughly studied in \cite{Grebenkov18b}.

\begin{figure}
\begin{center}
\includegraphics[width=88mm]{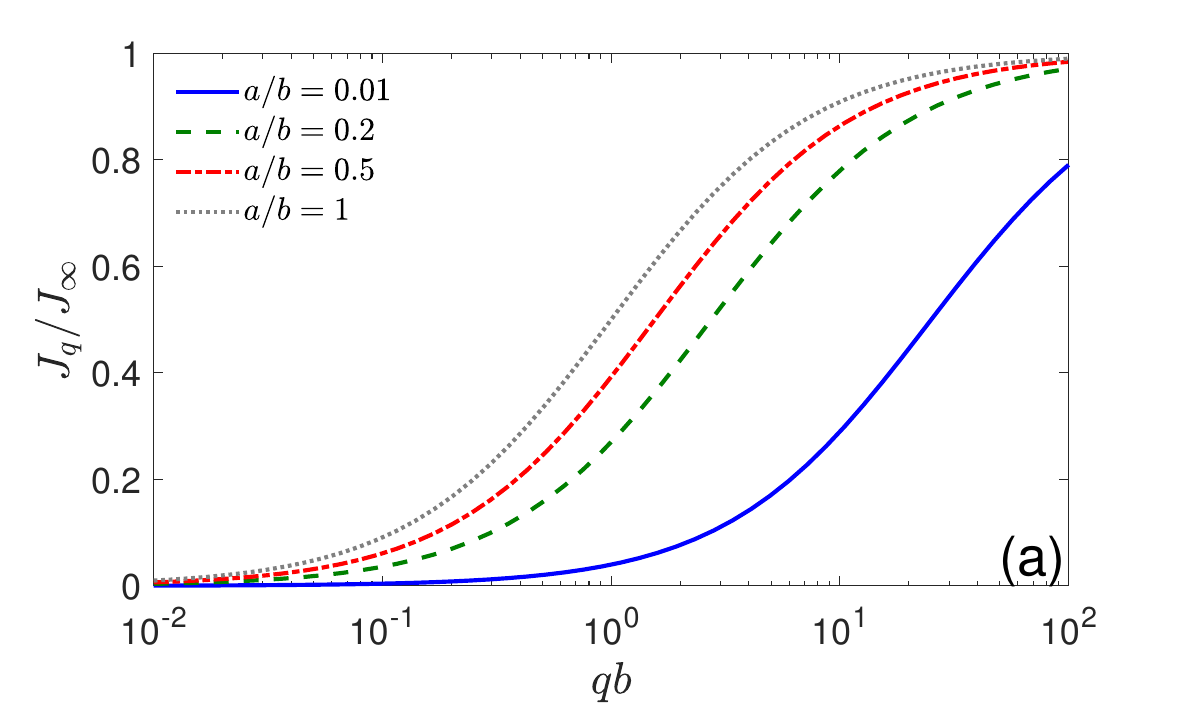} 
\includegraphics[width=88mm]{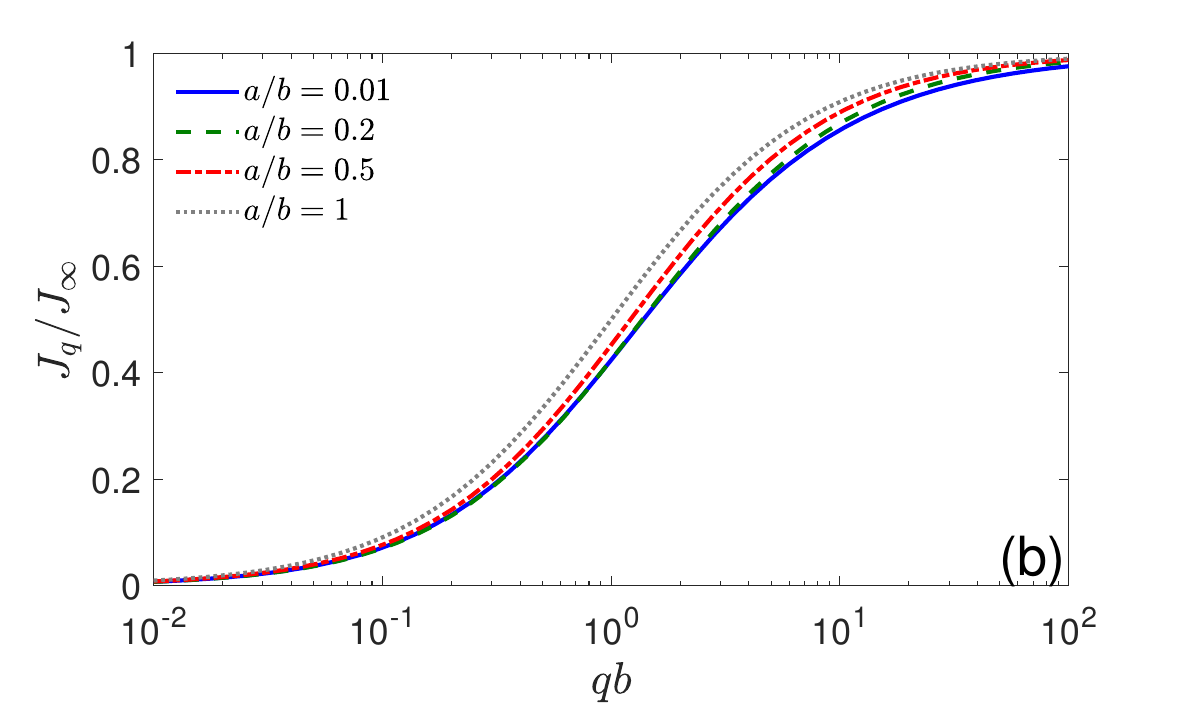} 
\includegraphics[width=88mm]{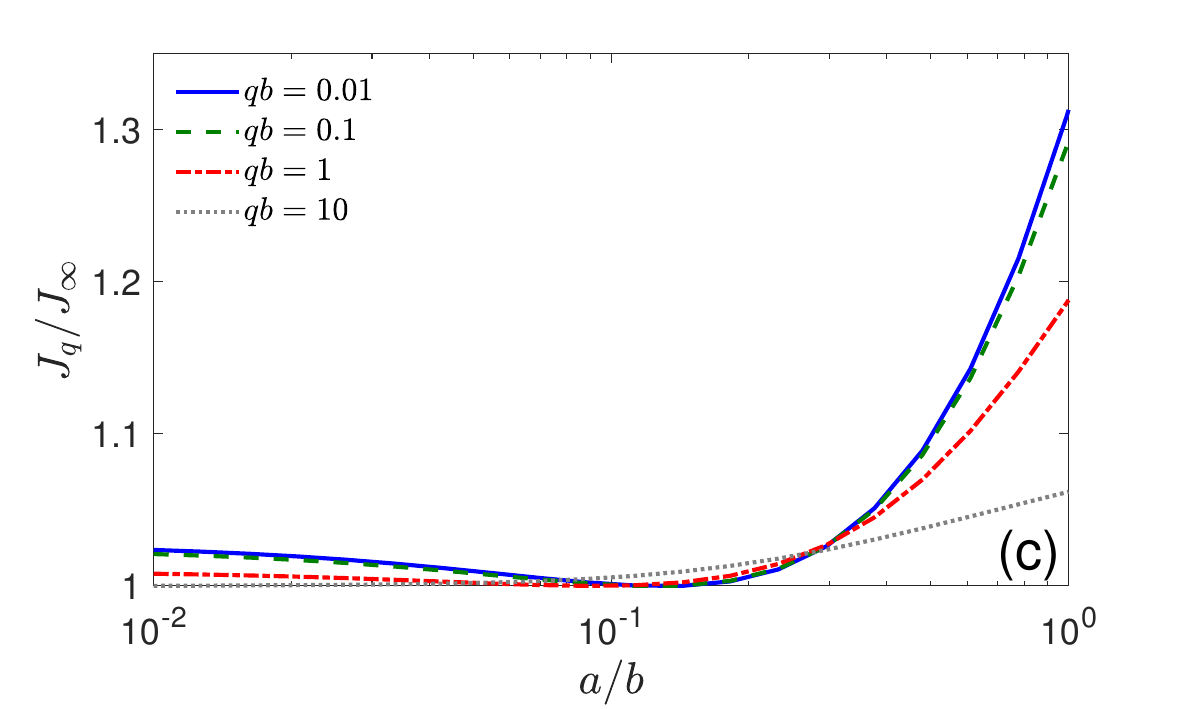} 
\end{center}
\caption{
The diffusive flux $J_q$ onto a partially reactive spheroidal boundary
with semi-axes $a$ and $b$, rescaled by $J_\infty$.  Panels {\bf (a)}
and {\bf (b)} present $J_q$ as a function of $qb$ for prolate and
oblate spheroids, respectively, whereas panel {\bf (c)} shows $J_q$ as
a function of $a/b$ for oblate spheroids.  For an easier comparison
between different curves, the ratio $J_q/J_\infty$ on panel {\bf (c)}
was rescaled by its minimum (so that all the minimal ratios are equal
to $1$).  The matrix $\tilde{\MM}$ in Eqs. (\ref{eq:Jq_prolate},
\ref{eq:Jq_oblate}) was truncated to the size $(\nmax+1)^2 \times
(\nmax+1)^2$, with $\nmax = 10$.  }
\label{fig:Jq}
\end{figure}

Figure \ref{fig:Jq} illustrates the behavior of the diffusive flux
$J_q$, rescaled by $J_\infty$, as a function of the reactivity
parameter $q$.  Expectedly, this ratio changes from $0$ at $q = 0$ (no
surface reaction) to $1$ at $q\to\infty$ (perfect surface reaction).
For prolate spheroids, their anisotropy reduces the diffusive flux
monotonously (curves are shifted downwards as $a/b$ decreases).  As
the spheroid gets thinner, its accessibility to Brownian motion is
reduced, and this effect is enhanced by decreasing the reactivity $q$
and thus requiring more and more returns to the target to realize a
successful reaction.  The situation is more subtle for oblate
spheroids.  Even in the limit $a = 0$, the disk remains accessible to
Brownian motion so that the effect of anisotropy onto the diffusive
flux is moderate (all shown curves are close to each other).
Curiously, the reduction of the diffusive flux by anisotropy is not
monotonous for oblate spheroids.  For instance, one can notice that
the curves for $a/b = 0.01$ and $a/b = 0.2$ cross each other.  This
non-monotonicity is illustrated on panel (c) which shows the ratio
$J_q/J_\infty$ as a function of $a/b$.  When $q$ is not too large,
this ratio exhibits a minimum at some intermediate aspect ratio that
depends on the reactivity parameter $q$.  From the mathematical point
of view, this behavior may be a consequence of the non-monotonous
dependence of eigenvalues $\mu_{mn}$ of the Dirichlet-to-Neumann
operator on the aspect ratio $a/b$, as illustrated in
Fig. \ref{fig:muk_oblate_ext}.  From the physical point of view, it
may result from an intricate interplay between the accessibility of
the surface controlled by its curvature (i.e., by $a/b$) and the
number of failed reaction attempts on it due to partial reactivity
(controlled by $qb$).  For instance, if the flux $J_q$ was not
rescaled by $J_\infty$, there would be no minimum.  Further
clarifications of this effect require a more detailed analysis of the
diffusive dynamics near curved, partially reactive boundaries.  We
stress, however, that this effect is moderate for oblate spheroids.

More generally, the Dirichlet-to-Neumann operator and its matrix
representation allow one to investigate the effect of spatially
heterogeneous reactivity when $q$ changes on the boundary
\cite{Grebenkov19}.  In addition, one can use the Steklov
eigenfunctions to compute the Green's function with the Robin boundary
condition.

\subsection{Statistics of boundary encounters}
\label{sec:encounters}

The Steklov eigenfunctions also play the central role in the
encounter-based approach to diffusion-controlled reactions
\cite{Grebenkov20}.  In this context, one usually considers a more
general setting, in which the Laplace equation (\ref{eq:Laplace}) is
replaced by the modified Helmholtz equation, $(p-D \Delta) u = 0$,
with $p \geq 0$.  The eigenfunctions of the related Steklov problem
determine the spectral expansion of the Laplace transform of the
encounter propagator $P(\x,\ell,t|\x_0)$, i.e., the joint probability
density of the position $\X_t$ and the boundary local time $\ell_t$ of
a particle that starts from a point $\x_0$ and diffuses in a domain
$\Omega$ with the reflecting boundary $\pa$ \cite{Grebenkov20}.  Here
the boundary local time $\ell_t$ can be understood as the (rescaled)
number of encounters between the diffusing particle and the boundary,
which is crucial for describing various phenomena occuring on the
boundary \cite{Grebenkov19c,Grebenkov20c,Grebenkov21a,Grebenkov22d}.
The encounter propagator allows one to access most common
characteristics of diffusion-controlled reactions, including the
conventional propagator, the survival probability, the first-passage
time probability density, the harmonic measure density, the diffusive
flux, etc.  For instance, one can investigate the first-crossing
time $\tau = \inf\{t > 0 ~:~ \ell_t > \ell\}$ of a threshold $\ell$ by
the boundary local time $\ell_t$.  In other words, the random variable
$\tau$ is the first time instance when the number of encounters
exceeds a given threshold.  The probability density of this random
variable,
\begin{equation}
U(\ell,t|\x_0) dt = \P_{\x_0}\{\tau \in (t,t+dt)\}  ,
\end{equation}
was shown to be expressed in terms of the integral of
$P(\x,\ell,t|\x_0)$ over the boundary $\pa$ (see details in
\cite{Grebenkov20,Grebenkov23}):
\begin{equation}
U(\ell,t|\x_0) = D \int\limits_{\pa} d\x \, P(\x,\ell,t|\x_0).
\end{equation}
For a bounded domain, any threshold is crossed with probability $1$
so that the probability density $U(\ell,t|\x_0)$ is normalized to $1$.
However, when a three-dimensional domain is unbounded, the particle
can escape to infinity and never return to the boundary $\pa$.  As a
consequence, the integral of $U(\ell,t|\x_0)$ over $t$ may be smaller
than $1$; in fact, this integral determines the probability of
crossing of a given threshold $\ell$:
\begin{equation}
\P_{\x_0}\{ \ell_\infty > \ell\} = \int\limits_0^\infty dt \, U(\ell,t|\x_0) ,
\end{equation}
which can be seen as the Laplace transform of $U(\ell,t|\x_0)$,
evaluated at $p = 0$.  One can therefore employ the general spectral
expansion for the Laplace transform of the encounter propagator
$P(\x,\ell,t|\x_0)$, derived in \cite{Grebenkov20}, to determine this
probability in terms of the Steklov eigenfunctions for the Laplace
equation (i.e., for $p = 0$) that we studied here:
\begin{equation}
\P_{\x_0}\{ \ell_\infty > \ell\} 
= \sum\limits_{k=0}^\infty \frac{[V_k(\x_0)]^*}{\|v_k\|_{L^2(\pa)}^2} e^{-\mu_k \ell} \int\limits_{\pa} d\x \, v_k(\x).
\end{equation}

\begin{figure}
\begin{center}
\includegraphics[width=88mm]{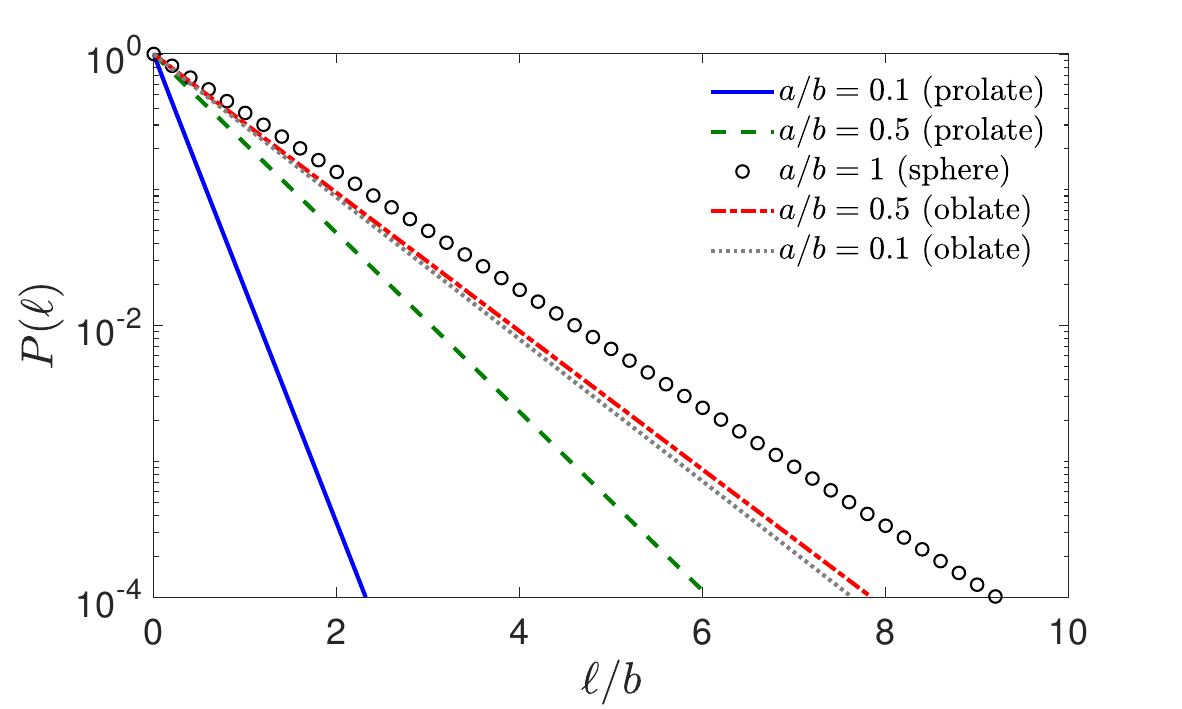} 
\end{center}
\caption{
The crossing probability $P(\ell)$ from Eq. (\ref{eq:Pell_def}) as a
function of the threshold $\ell$ for prolate and oblate spheroids with
semi-axes $a$ and $b$.  The starting point is uniformly distributed on
the boundary.  The matrix $\tilde{\MM}$ in Eq. (\ref{eq:Pell_M}) was
truncated to the size $(\nmax+1)^2 \times (\nmax+1)^2$, with $\nmax =
10$.  }
\label{fig:Pell}
\end{figure}

At $\ell = 0$, this is simply the probability of hitting the boundary
(before escaping to infinity).  At large $\ell$, this probability
decays exponentially, and the rate of this decay is given by the
smallest eigenvalue $\mu_0$.  When prolate spheroids get thinner
($a\to 0$), the eigenvalue $\mu_0$ diverges according to
Eq. (\ref{eq:mu_prolate_a0}), so that the crossing probability
$\P_{\x_0}\{ \ell_\infty > \ell\}$ decays faster with $\ell$.  Indeed,
it is hard for Brownian motion to access thin prolate spheroids, and
it is therefore unlikely to experience many encounters with it (i.e.,
to get large $\ell_\infty$).  In turn, the eigenvalue $\mu_0$
approaches a constant value in the opposite limit of thin oblate
spheroids.  This is consistent with the fact that even a disk ($a =
0$) remains accessible to Brownian motion in three dimensions.

For illustrative purposes, we consider the starting point to be
uniformly distributed over the spheroidal boundary and evaluate the
the associated crossing probability:
\begin{align}   \label{eq:Pell_def}
P(\ell) & = \frac{1}{|\pa|} \int\limits_{\pa} d\x_0 \, \P_{\x_0}\{ \ell_\infty > \ell\}  \\  \nonumber
& = \frac{1}{|\pa|} \sum\limits_{k=0}^\infty e^{-\mu_k \ell} \frac{|(1,v_k)|^2_{L^2(\pa)}}{\|v_k\|^2_{L^2(\pa)}} \,.
\end{align}
Comparing this expression with the spectral expansion of the total
flux $J_q$ (see the third line in Eq. (\ref{eq:Jq_prolate})), one can
realize that
\begin{equation}
\frac{J_q}{Dc_0 q |\pa|} = \int\limits_0^\infty d\ell \, e^{-q\ell} \,\rho(\ell),
\end{equation}
where $\rho(\ell) = -\partial_\ell P(\ell)$ is the probability density
of the random variable $\ell_\infty$.  As a consequence, we can
formally invert the matrix representation (\ref{eq:Jq_prolate}) to get
\begin{equation}  \label{eq:Pell_M}
P(\ell) = \frac{\C}{|\pa|} \bigl[ e^{-\ell \tilde{\MM}} /\tilde{\MM} \bigr]_{00,00} 
\end{equation}
for both prolate and oblate spheroids, with $\C$ being their harmonic
capacity given by Eqs. (\ref{eq:Charm_prolate}, \ref{eq:Charm_oblate}).

Figure \ref{fig:Pell} shows the behavior of the crossing probability
$P(\ell)$ for prolate and oblate spheroids, as well as for the sphere
of radius $b$, for which $P(\ell) = e^{-\ell/b}$.  The slowest decay
of $P(\ell)$ corresponds to the sphere, which is the most accessible
to Brownian motion.  As $a/b$ goes to $0$, the decay is getting faster
and faster for prolate spheroids.  In turn, for oblate spheroids, the
crossing probability $P(\ell)$ approaches that of a disk.  The role of
the starting point $\x_0$ and other properties of the boundary local
time $\ell_\infty$ can be investigated using our results.

\section{Conclusions}
\label{sec:conclusion}

In this paper, we investigated the spectral properties of the
Dirichlet-to-Neumann operator $\M$ and the related Steklov problem for
anisotropic domains in three dimensions.  Using prolate and oblate
spheroidal coordinates, we derived a matrix representation of $\M$ on
the basis of spherical harmonics for both exterior and interior
problems.  Its diagonalization allowed us to access the eigenvalues
$\mu_{mn}$ and eigenfunctions $v_{mn}$ of $\M$, as well as the Steklov
eigenfunctions $V_{mn}$.  These eigenfunctions inherited the
symmetries of the considered domains; in particular, $V_{mn}$ depend
on the angle $\phi$ via the factor $e^{im\phi}$, and are symmetric
(resp., antisymmetric) with respect to the horizontal plane $z = 0$
when $m+n$ is even (resp., odd).  As a consequence, they are also the
eigenfunctions of the mixed Steklov-Neumann (resp., Steklov-Dirichlet)
problems in the upper half-space.  We also described recurrence
relations to speed up the numerical construction of truncated
matrices.  While the interior spectral problem in a bounded domain
could alternatively be solved by other numerical methods
\cite{Andreev04,Yang09,Bi11,Li11,Li13,Xie14,Bi16,Bogosel16,Akhmetgaliyev17,Alhejaili19,Chen20},
exterior spectral problems in unbounded domains are in general more
difficult to address by conventional tools.

We discussed the impact of domain anisotropy onto the behavior of
eigenvalues and eigenfunctions.  In particular, we showed how the
eigenvalues of $\M$ for the exterior of a prolate spheroid diverge in
the limit $a\to 0$ (the exterior of a needle); in turn, in the
opposite limit of thin (disk-like) oblate spheroids, the eigenvalues
reach finite values.  In this limit, we also got complementary
insights onto a classical mixed Dirichlet-Neumann problem in the
half-space.

Apart from their own fundamental interest in spectral geometry, the
Steklov eigenfunctions offer flexible meshless representations for
solutions of interior and exterior boundary value problems.  In the
context of diffusion-controlled reactions, these eigenfunctions allow
one to incorporate the effect of partial reactivity, e.g., to compute
the steady-state concentration of particles that react on the
spheroidal boundary.  In particular, we showed how the total diffusive
flux (i.e., the overall reaction rate) depends on the anisotropy of
the target.  Another application from statistical physics concerns the
statistics of encounters between a diffusing particle and the
spheroidal target.  The Steklov eigenfunctions determine the crossing
probability $\P_{\x_0}\{ \ell_\infty > \ell\}$, i.e., the probability
that the (rescaled) number of encounters, $\ell_\infty$, exceeds a
given threshold $\ell$, before the particle escapes to infinity.  The
crossing probability generalizes the hitting probability of the target
(the latter corresponding to $\ell = 0$).  In particular, we showed
that the smallest eigenvalue $\mu_{00}$ of the Dirichlet-to-Neumann
operator determines the exponential decay of the crossing probability
at large thresholds $\ell$.

In summary, this study brings novel insights onto the spectral
properties of the Dirichlet-to-Neumann operator for exterior domains,
with the special emphasis on anisotropy.  Even though the derived
eigenvalues and eigenfunctions are not as explicit as in the case of
the exterior of a ball, the matrix representation offers an efficient
numerical computation and allows for getting asymptotic results.  One
straightforward extension of this study concerns diffusion across a
semi-permeable spheroidal boundary, which is a model of diffusive
exchange in anisotropic living cells or tissues
\cite{Richter74,Grimes18,Gadzinowski21}.  The exchange between
interior and exterior compartments is usually described by a
transmission boundary condition that can be reformulated in terms of
two Dirichlet-to-Neumann operators for the interior and exterior
problems.  Moreover, an extension of the encounter-based approach to
semi-permeable membranes allows one to treat much more sophisticated
exchange mechanisms
\cite{Bressloff22a,Bressloff22b,Bressloff22c,Bressloff22d}.
Another promising direction is related to the Dirichlet-to-Neumann
operator for the (modified) Helmholtz equation.  As briefly discussed
in Sec. \ref{sec:encounters}, the related Steklov eigenfunctions
determine the encounter propagator and thus most diffusion-reaction
characteristics in this system.  While one could employ prolate/oblate
spheroidal wave functions, their analysis is more sophisticated and
less explicit.  Alternatively, one can consider other domains, in
which the Laplace operator admits separation of variables in
appropriate curvilinear coordinates (e.g., a torus).  A systematic
investigation of the Dirichlet-to-Neumann operator in such relatively
simple shapes can help to uncover the intricate relations between the
geometry and the spectral properties of this operator.

\begin{acknowledgments}
The author thanks A. Chaigneau for his help on the numerical
validation of the proposed method by its comparison to an alternative
construction of the Dirichlet-to-Neumann operator via a finite-element
method.  A partial support by the Alexander von Humboldt Foundation
within a Bessel Prize award is acknowledged.
\end{acknowledgments}

\appendix
\section{Interior problems}
\label{sec:int}

\begin{figure}[t!]
\begin{center}
\includegraphics[width=88mm]{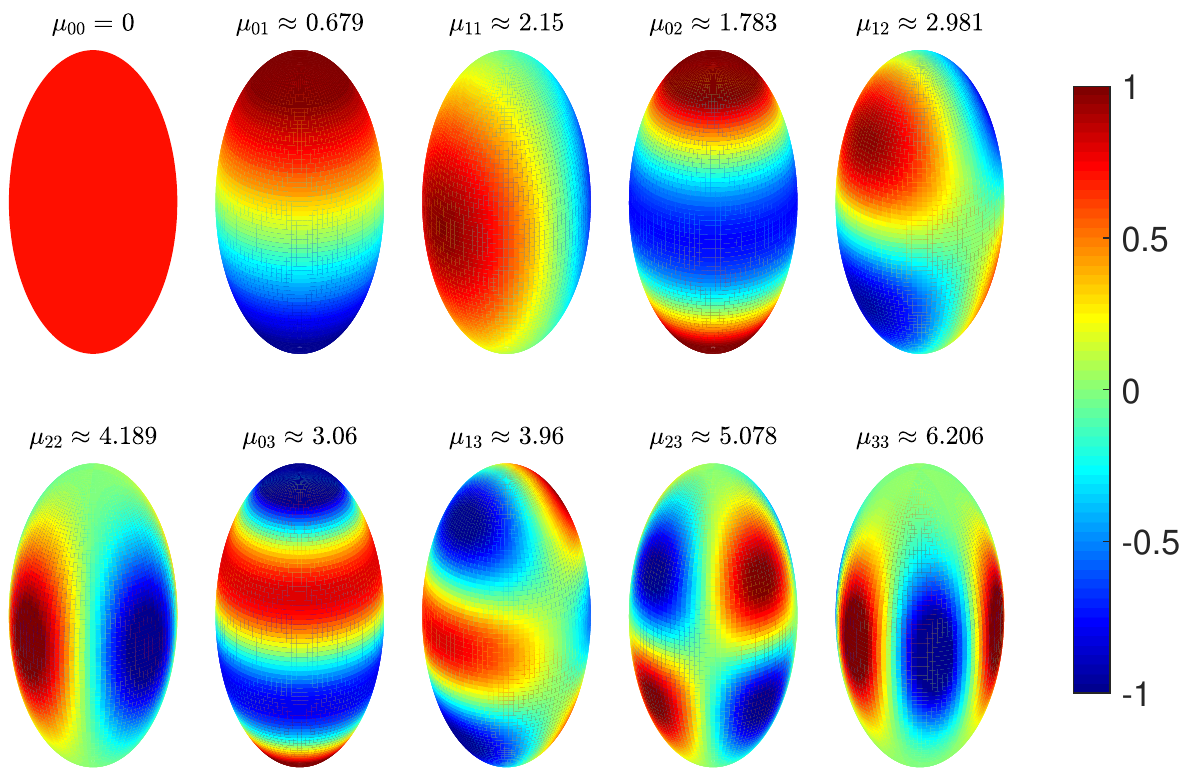} 
\end{center}
\caption{
Several eigenfunctions $v_{mn}$ of the Dirichlet-to-Neumann operator
$\M$ for the interior of the prolate spheroid with semi-axes $a = 0.5$
and $b = 1$.  The associated eigenvalues are shown on the top.  The
truncation order is $\nmax = 10$.  }
\label{fig:vn_prolateInt}
\end{figure}

While the main text is focused on exterior problems, the same approach
is valid for the interior spectral problems when the Laplace equation
has to solved inside a bounded prolate or oblate spheroid.  In fact,
the coefficients $F_{n,n'}^m(z)$ and $\bar{F}_{n,n'}^m(z)$ remain
unchanged, and the difference between exterior and interior settings
is only manifested in the coefficients $c_{mn}$:

(i) for prolate spheroids, it is sufficient to replace $Q_n^m(z)$ by
$P_n^m(z)$ in Eq. (\ref{eq:cmn_prolate}), as well as the sign:
\begin{equation}  \label{eq:cmn_prolate_int}
c_{mn} = \frac{\sinh \alpha_0 \, P^{'m}_n(\cosh \alpha_0)}{a_E P_n^m(\cosh\alpha_0)} \,.
\end{equation}

(ii) for oblate spheroids, it is sufficient to replace $Q_n^m(z)$ by
$P_n^m(z)$ in Eq. (\ref{eq:cmn_oblate}), as well as the sign:
\begin{equation}  \label{eq:cmn_oblate_int}
c_{mn} = \frac{i\cosh \alpha_0 \, P^{'m}_n(i\sinh \alpha_0)}{a_E P_n^m(i\sinh\alpha_0)} \,.
\end{equation}
In both cases, as $P'_0 = 0$ and thus $c_{00} = 0$, one can easily
check that the first eigenvalue is zero, $\mu_{00} = 0$, as it should
be.

\begin{figure}
\begin{center}
\includegraphics[width=88mm]{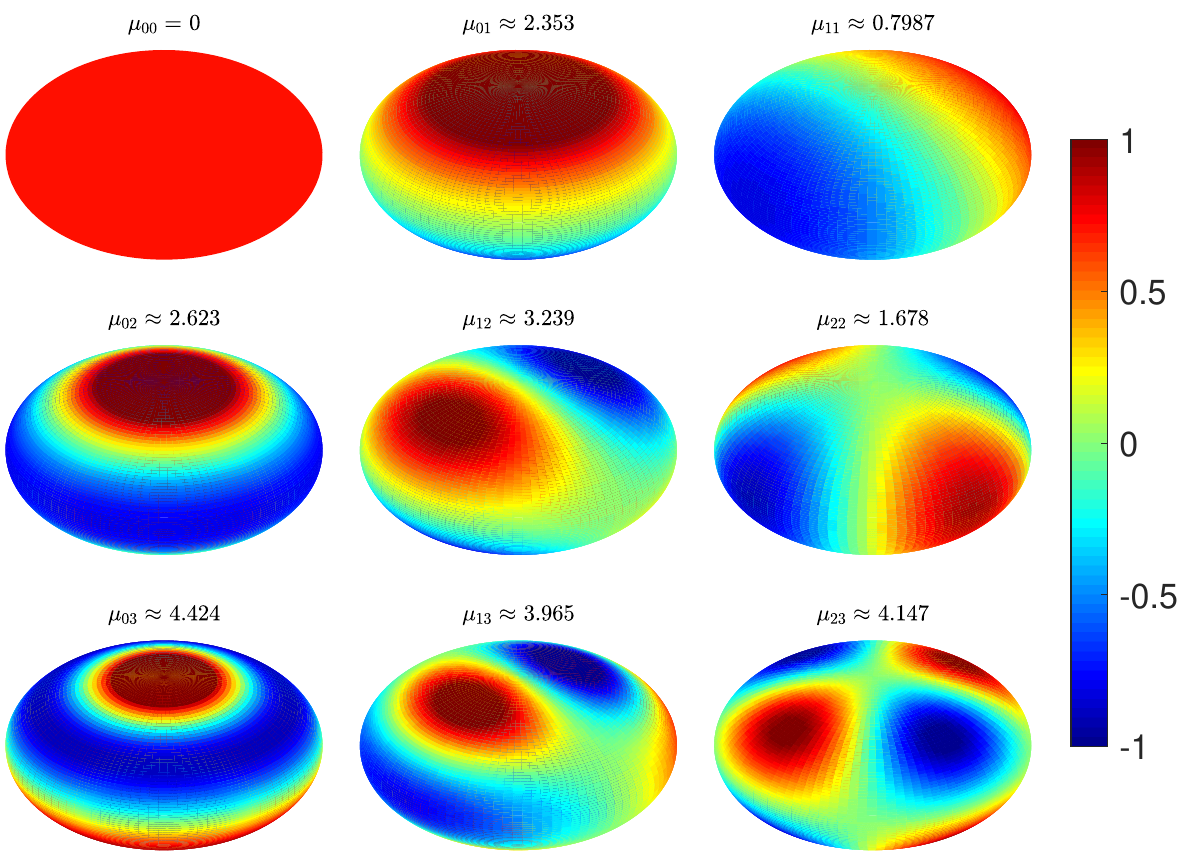} 
\end{center}
\caption{
Several eigenfunctions $v_{mn}$ of the Dirichlet-to-Neumann operator
$\M$ for the interior of the oblate spheroid with semi-axes $a = 0.5$
and $b = 1$.  The associated eigenvalues are shown on the top.  The
truncation order is $\nmax = 10$.  }
\label{fig:vn_oblateInt}
\end{figure}

Figures \ref{fig:vn_prolateInt} and \ref{fig:vn_oblateInt} illustrate
several eigenfunctions $v_{mn}$ for the interior prolate/oblate
spheroid with semi-axes $a = 0.5$ and $b = 1$.  The associated
eigenvalues $\mu_{mn}$ are shown on the top of each panel.  
Expectedly, a constant eigenfunction $v_{00}$ corresponds to $\mu_{00}
= 0$.  Other eigenfunctions also resemble the spherical harmonics.

One can also consider more sophisticated domains confined between two
confocal prolate (oblate) spheroids.  These domains are characterized
as $\alpha_1 < \alpha < \alpha_2$, where $\alpha_1 =
\tanh^{-1}(a/b)$ and $\alpha_2 = \tanh^{-1}(A/B)$, with $a,b$ and
$A,B$ being the semi-axes of the inner and outer boundaries such that
$a_E = \sqrt{b^2 - a^2} = \sqrt{B^2 - A^2}$ is half of the focal
distance for both boundaries.  In this case, a general solution of the
Laplace equation involves both $P_n^m(z)$ and $Q_n^m(z)$ but the
structure of the matrix representation is very similar.  We just
mention two settings, in which either Dirichlet or Neumann boundary
condition is imposed on the outer boundary, whereas the inner boundary
has the Steklov condition (the opposite setting can be easily
obtained).

(i) For prolate spheroids, the matrix $\MM$ is still given by
Eq. (\ref{eq:M_prolate}), in which $\alpha_0$ is replaced by
$\alpha_1$, and the coefficients $c_{mn}$ depend on the boundary
condition on the outer spheroid:
\begin{widetext}
\begin{align}
& c_{mn} = - \frac{\sinh\alpha_1}{a_E} \, \frac{P_n^m(\cosh\alpha_2) Q_n^{'m}(\cosh\alpha_1) - Q_n^m(\cosh\alpha_2) P_n^{'m}(\cosh\alpha_1)}
{P_n^m(\cosh\alpha_2) Q_n^{m}(\cosh\alpha_1) - Q_n^m(\cosh\alpha_2) P_n^{m}(\cosh\alpha_1)}  \qquad \textrm{(Dirichlet),} \\
& c_{mn} = - \frac{\sinh\alpha_1}{a_E} \, \frac{P_n^{'m}(\cosh\alpha_2) Q_n^{'m}(\cosh\alpha_1) - Q_n^{'m}(\cosh\alpha_2) P_n^{'m}(\cosh\alpha_1)}
{P_n^{'m}(\cosh\alpha_2) Q_n^{m}(\cosh\alpha_1) - Q_n^{'m}(\cosh\alpha_2) P_n^{m}(\cosh\alpha_1)}  \qquad \textrm{(Neumann).}
\end{align}
\end{widetext}

(ii) For oblate spheroids, the matrix $\MM$ is still given by
Eq. (\ref{eq:M_oblate}), in which $\alpha_0$ is replaced by
$\alpha_1$, and the coefficients $c_{mn}$ depend on the boundary
condition on the outer spheroid:
\begin{widetext}
\begin{align}
& c_{mn} = \frac{\cosh\alpha_1}{ia_E} \, \frac{P_n^m(i\sinh\alpha_2) Q_n^{'m}(i\sinh\alpha_1) - Q_n^m(i\sinh\alpha_2) P_n^{'m}(i\sinh\alpha_1)}
{P_n^m(i\sinh\alpha_2) Q_n^{m}(i\sinh\alpha_1) - Q_n^m(i\sinh\alpha_2) P_n^{m}(i\sinh\alpha_1)}  \qquad \textrm{(Dirichlet),} \\
& c_{mn} = \frac{\cosh\alpha_1}{ia_E} \, \frac{P_n^{'m}(i\sinh\alpha_2) Q_n^{'m}(i\sinh\alpha_1) 
- Q_n^{'m}(i\sinh\alpha_2) P_n^{'m}(i\sinh\alpha_1)}
{P_n^{'m}(i\sinh\alpha_2) Q_n^{m}(i\sinh\alpha_1) - Q_n^{'m}(i\sinh\alpha_2) P_n^{m}(i\sinh\alpha_1)}  \qquad \textrm{(Neumann).}
\end{align}
\end{widetext}

\section{Numerical computation}
\label{sec:numerics}

In this section, we discuss a practical implementation of the proposed
method.  The construction of the (truncated) matrix $\MM$ requires
computing the integral (\ref{eq:Fmnn}) and the coefficients $c_{mn}$,
both involving associated Legendre functions $P_n^m(x)$ and
$Q_n^m(x)$.  For this reason, we briefly summarize the main steps in
the evaluation of these functions and related integrals.

\subsection{Associated Legendre functions}

The definition and basic properties of associated Legendre functions
$P_n^m(x)$ and $Q_n^m(x)$ can be found in many textbooks, e.g., in
\cite{Abramowitz} (chapter 8).  As we focus on integer
indices $n = 0,1,2,\ldots$ and $m = -n,-n+1,\ldots,n$, the numerical
computation of $P_n^m(x)$ is fairly standard in the common range $-1
\leq x \leq 1$ (see, e.g., the built-in function \verb|legendre| in
matlab).  However, the computation of the coefficients $c_{mn}$
requires evaluating both $P_n^m(z)$ and $Q_n^m(z)$ at $z = \cosh\alpha
> 1$ or $z = i\sinh \alpha$.  While $P_n^m(z)$ with even $m$ are just
polynomials and thus can be immediately evaluated at any complex
number, the computation is more subtle for other associated Legendre
functions that involve square roots and logarithms and thus require
cuts in the complex plane.  Throughout this section, we follow
Ref. \cite{Abramowitz} and distinguish two conventions by writing the
argument as $x$ for $x \in (-1,1)$, or as $z$ for $z \notin [-1,1]$.
For instance, two conventional definitions read for $m\geq 0$ as:
\begin{subequations}
\begin{align}
P_n^m(x) & = (-1)^m (1-x^2)^{m/2} \frac{d^m}{dx^m} P_n(x), \\
P_n^m(z) & = (z^2-1)^{m/2} \frac{d^m}{dz^m} P_n(z), 
\end{align}
\end{subequations}
and
\begin{subequations}
\begin{align}
Q_n^m(x) & = (-1)^m (1-x^2)^{m/2} \frac{d^m}{dx^m} Q_n(x), \\  \label{eq:Qnm_def}
Q_n^m(z) & = (z^2-1)^{m/2} \frac{d^m}{dz^m} Q_n(z), 
\end{align}
\end{subequations}
where $P_n(z)$ are the Legendre polynomials and $Q_n(z)$ are the
Legendre functions of the second kind (see below).  In the following,
we focus on $P_n^m(z)$ and $Q_n^m(z)$ and briefly describe their
numerical computation via the standard recurrence formulas for
completeness.  As these formulas are identical for both $P_n^m(z)$ and
$Q_n^m(z)$, we consider the latter functions and then just mention
changes for $P_n^m(z)$.

In the first step, one can compute the Legendre functions $Q_n(z)$ up
to the desired order $n$ via the recurrence relation,
\begin{equation}
Q_n(z) = \frac{2n-1}{n} z Q_{n-1}(z) - \frac{n-1}{n} Q_{n-2}(z),
\end{equation}
starting from
\begin{equation}  \label{eq:Q0}
Q_0(z) = \frac12 \ln \biggl(\frac{z+1}{z-1}\biggr) , \quad Q_1(z) = z Q_0(z) - 1. 
\end{equation}
From Eq. (\ref{eq:Qnm_def}), one can also evaluate 
\begin{align}
& Q_n^1(z) = \frac{1}{n \sqrt{z^2-1}} \biggl(n(2n-1) z Q_n(z) - \\  \nonumber
& [(2n-1)(n-1)z^2 + n^2]Q_{n-1}(z) + (n-1)^2 z Q_{n-2}(z)\biggr)
\end{align}
for $n \geq 2$, where we used another recurrence relation
\begin{equation}  \label{eq:dQn}
\frac{d}{dz} Q_n(z) = \frac{n+1}{z^2-1} \bigl(Q_{n+1}(z) - z Q_n(z)\bigr).
\end{equation}
One also has
\begin{equation}  \label{eq:Q11}
Q_1^1(z) = \sqrt{z^2-1}\, Q_0(z) - \frac{z}{\sqrt{z^2-1}} \,.
\end{equation}

Since we know $Q_n^0(z) = Q_n(z)$ and $Q_n^1(z)$, the other functions
$Q_n^m(z)$ can be found via the recurrence relation:
\begin{equation}
Q_n^{m-1}(z) = \frac{2mz (z^2-1)^{-\frac12} Q_n^m(z) + Q_n^{m+1}(z)}{(n+m)(n-m+1)} 
\end{equation}
for $m = -n+2,-n+3,\ldots,-1,0$, or 
\begin{equation}
Q_n^{m+1}(z) = (n+m)(n-m+1) Q_n^{m-1}(z) - \frac{2mz Q_n^m(z)}{(z^2-1)^{\frac12}}  
\end{equation}
for $m = 1,2,\ldots,n$.  Note that the derivative of $Q_n^m(z)$ can be
found via either of two equivalent relations
\begin{align}  \nonumber
\frac{dQ_n^m(z)}{dz} &= \frac{1}{z^2-1} \biggl[(n+m)(n-m+1) \sqrt{z^2-1}\, Q_n^{m-1}(z) \\
& - mz Q_n^m(z)\biggr], \\
&= \frac{1}{z^2-1} \biggl[\sqrt{z^2-1}\, Q_n^{m+1}(z) + mz Q_n^m(z)\biggr].
\end{align}

The computation of $P_n^m(z)$ relies on the same recurrence relations,
except for their initiation, namely, Eqs. (\ref{eq:Q0}, \ref{eq:Q11})
are replaced by
\begin{equation}  \label{eq:P0}
P_0(z) = 1 , \quad P_1(z) = z,  \quad  P_1^1(z) = \sqrt{z^2-1}.
\end{equation}

\subsection{Alternative matrix representation}
\label{sec:alternative}

In the limit $\alpha_0 \to 0$, the coefficients
$\bar{F}_{n,n'}^m(\sinh\alpha_0)$ given by Eq. (\ref{eq:Fmn_oblate})
logarithmically diverge that makes their numerical computation more
subtle.  It is therefore convenient to get an alternative matrix
representation that can be constructed even at $\alpha_0 = 0$.  For
this purpose, we apply the operator $\M$ to Eq. (\ref{eq:vmn_oblate})
that yields
\begin{align*}
& \mu_{mn} \sum\limits_{n'=0}^\infty [\VV_{mn}]_{mn'} \bar{Y}_{mn'} = \mu_{mn} v_{mn} = \M v_{mn} \\
& = \sum\limits_{n'=0}^\infty [\VV_{mn}]_{mn'} c_{mn'} \frac{\bar{Y}_{mn'}}{\sqrt{\cosh^2\alpha_0 - \cos^2\theta}} \,,
\end{align*}
where we wrote explicitly the action of $\M$ onto a spherical harmonic
$\bar{Y}_{mn'}$ via Eq. (\ref{eq:Mu_oblate}).  Multiplying both sides
of this relation by $\sqrt{\cosh^2\alpha_0 - \cos^2\theta} \,
\bar{Y}^*_{mn_0} \cos\theta$ and integrating over $\theta$ and $\phi$,
we get
\begin{equation} \label{eq:auxil12}
\mu_{mn} \sum\limits_{n'=0}^\infty [\VV_{mn}]_{mn'} \bar{G}_{mn',mn_0}(\sinh\alpha_0) = [\VV_{mn}]_{mn_0} c_{mn_0}  ,
\end{equation}
where we used the orthogonality of spherical harmonics, and defined
\begin{align} \nonumber
& \bar{G}_{mn,m'n'}(z) = \int\limits_0^{2\pi} d\phi 
\int\limits_{-\pi/2}^{\pi/2} d\theta \, \cos\theta \, \bar{Y}_{mn}(\theta,\phi) \\  \nonumber
& \times \sqrt{z^2 + \sin^2\theta} \,  \bar{Y}_{m'n'}^*(\theta,\phi) \\  \label{eq:Gmn0}
& = 2\pi \delta_{m,m'} a_{mn} a_{m'n'} G_{n,n'}^m(z) ,
\end{align}
with
\begin{equation}  \label{eq:Gmn}
G_{n,n'}^m(z) = \int\limits_{-1}^{1} dx \, P_{n}^m(x) P_{n'}^m(x)  \sqrt{z^2 + x^2} .
\end{equation}
These coefficients resemble $\bar{F}_{n,n'}^m(z)$ from
Eq. (\ref{eq:Fmn_oblate}), except that the square root $\sqrt{z^2 +
x^2}$ stands in the numerator and thus eliminates the divergence at $z
= 0$.  Finally, Eq. (\ref{eq:auxil12}) can be written as an eigenvalue
problem
\begin{equation} \label{eq:auxil13}
\sum\limits_{n'=0}^\infty [\VV_{mn}]_{mn'}  \frac{\bar{G}_{mn',mn_0}(\sinh\alpha_0)}{c_{mn_0}} = \frac{1}{\mu_{mn}}  [\VV_{mn}]_{mn_0}  .
\end{equation}
This relation implies that $\VV_{mn}$ is a {\it left} eigenvector of
the matrix with the elements
$\bar{G}_{mn',mn_0}(\sinh\alpha_0)/c_{mn_0}$, whereas $1/\mu_{mn}$ is
the associated eigenvalue.  One sees that this is a matrix
representation of $\M^{-1}$, the inverse of the Dirichlet-to-Neumann
operator, which is called the Neumann-to-Dirichlet operator.  The
numerical advantage of this matrix representation is that the elements
$\bar{G}_{n,n'}^m(z)$ do not diverge as $z\to 0$.  In particular, one
can compute the matrix elements $\bar{G}_{mn',mn_0}(0)/c_{mn_0}$ and
then diagonalize the truncated matrix.

\subsection{Recurrence formulas for the integral}

Now we describe the recurrence formulas for computing the elements of
the matrix $\MM$.  These formulas are actually valid in a more general
case:
\begin{equation}  \label{eq:A_def}
\AA_{n,n'}^m = \int\limits_{-1}^1 dx \, f(x^2)\, P_n^m(x) \, P_{n'}^m(x),
\end{equation}
where $f(x)$ is a given integrable function.  Due to the symmetry
relation (\ref{eq:ALegendre_symm}), these integrals vanish when $n+n'$
is odd.  They also are zero if $|m| > n$ or $|m| > n'$.  The remaining
nonzero values can be found recursively by using the recurrence
relation for associated Legendre polynomials:
\begin{equation}
(n-m+1)P_{n+1}^m(x) = (2n+1) x P_{n}^m(x) - (n+m) P_{n-1}^m(x).
\end{equation}
Applying this relation to both $P_n^m(x)$ and $P_{n'}^m(x)$ in
Eq. (\ref{eq:A_def}), we get for even $n+n'$:
\begin{widetext}
\begin{align}  \nonumber
\AA_{n,n'}^m & = \int\limits_{-1}^1 dx \, f(x^2) \, P_n^m(x) \frac{(2n'-1)xP_{n'-1}^m(x) - (n'-1+m)P_{n'-2}^m(x)}{n'-m} \\  \nonumber
& = \int\limits_{-1}^1 dx \, f(x^2) \biggl[\frac{(n-m+1)P_{n+1}^m(x) + (n+m) P_{n-1}^m(x)}{2n+1}
(2n'-1) P_{n'-1}^m(x) - \frac{n'-1+m}{n'-m} P_n^m(x) P_{n'-2}^m \biggr] \\  \label{eq:Amnp}
& = \frac{2n'-1}{(2n+1)(n'-m)} \biggl((n+1-m)\AA_{n+1,n'-1}^m  
 + (n+m) \AA_{n-1,n'-1}^m\biggr)  - \frac{n'-1+m}{n'-m} \AA_{n,n'-2}^m .
\end{align}
\end{widetext}
As a consequence, if $\AA_{n,0}^m$ and $\AA_{n,1}^m$ are known, one
can evaluate the remaining elements.

For any $m > 0$, one has to evaluate
\begin{align*}
\AA_{n,m}^m & = 2 \int\limits_0^1 dx \, f(x^2)\, P_n^m(x) \, \underbrace{(-1)^m (2m-1)!! \,(1-x^2)^{m/2}}_{=P_m^m(x)} \\
& = -(2m-1) 2 \int\limits_0^1 dx \, f(x^2)  \, P_{m-1}^{m-1}(x) \, \sqrt{1-x^2} P_n^m(x) \\
& = \frac{2m-1}{2n+1} \biggl((n+m-1)(n+m) \AA_{n-1,m-1}^{m-1} \\
& - (n-m+1)(n-m+2) \AA_{n+1,m-1}^{m-1} \biggr)
\end{align*}
for even $m+n$, where we used another recurrence relation to express
$\sqrt{1-x^2} P_n^m(x)$ in terms of $P_{n\pm 1}^{m-1}(x)$.  Similarly,
one needs
\begin{align*}
\AA_{n,m+1}^m & = \frac{2m+1}{2n+1} \biggl((n+m-1)(n+m) \AA_{n-1,m}^{m-1} \\
& - (n-m+1)(n-m+2) \AA_{n+1,m}^{m-1} \biggr)
\end{align*}
for odd $m+n$.  In other words, once the elements $\AA_{n,n'}^{m-1}$
are constructed, one can find the elements $\AA_{n,n'}^m$.  
As a consequence, if one knows the elements $\AA_{n,0}^0$ and
$\AA_{n,1}^0$, one can first construct all the elements $\AA_{n,n'}^0$
by using recurrence relations (\ref{eq:Amnp}), and then progressively
get the elements $\AA_{n,n'}^m$.  We stress that this procedure does
not depend on the function $f$, which determines only the
initialization step, i.e., the elements $\AA_{n,0}^0$ and
$\AA_{n,1}^0$.

For the integrals $\bar{F}_{n,n'}^m(z)$ from
Eq. (\ref{eq:Fmn_oblate}), one has $f(y) = (z^2+y)^{-1/2}$, for which
we have
\begin{subequations}
\begin{align}
\AAF_{0,0}^0(z) & = \ln \left(\frac{\sqrt{z^2+1}+1}{\sqrt{z^2+1}-1}\right),  \\
\AAF_{1,1}^0(z) & = \sqrt{z^2+1} - \frac{z^2}{2} \ln \left(\frac{\sqrt{z^2+1}+1}{\sqrt{z^2+1}-1}\right).
\end{align}
\end{subequations}
Using the recurrence relations
\begin{equation}
P_n(x) = \frac{2n-1}{n} x P_{n-1}(x) - \frac{n-1}{n} P_{n-2}(x)
\end{equation}
for Legendre polynomials $P_n(x)$ and integrating by parts, one gets
two sets of relations
\begin{subequations}  
\begin{align}  \label{eq:An10}
\AAF_{n+1,0}^0(z) & = \frac{2n+1}{n+1} \AAF_{n,1}^0(z) - \frac{n}{n+1} \AAF_{n-1,0}^0(z) , \\
\AAF_{n+1,1}^0(z) & = - \frac{n-1}{n+2} \AAF_{n-1,1}^0(z) - \frac{2n+1}{n+2} z^2 \AAF_{n,0}^0(z) .
\end{align}
\end{subequations}
These relations determine all $\AAF_{n,0}^0(z)$ and $\AAF_{n,1}^0(z)$,
from which one can construct recursively all $\AAF_{n,n'}^m(z)$, as
described above.  According to Eq. (\ref{eq:barFmn_Fmn}), this
construction can also be applied to get the elements $F_{n,n'}^m(z)$.

For the integrals $\tilde{G}_{n,n'}^m(z)$ from Eq. (\ref{eq:Gmn}), one
has $f(y) = (z^2+y)^{1/2}$, for which we have
\begin{subequations}
\begin{align}
\AAG_{0,0}^0(z) & = \sqrt{z^2+1} + \frac{z^2}{2} \ln \left(\frac{\sqrt{z^2+1}+1}{\sqrt{z^2+1}-1}\right), \\
\AAG_{1,1}^0(z) & = \frac{(z^2+2) \sqrt{z^2+1}}{4} - \frac{z^4}{8} \ln \left(\frac{\sqrt{z^2+1}+1}{\sqrt{z^2+1}-1}\right).
\end{align}
\end{subequations}
Using the recurrence relations and integrating by parts, one retrieves
Eq. (\ref{eq:An10}), which is completed by
\begin{equation}
\AAG_{n+1,1}^0(z) = - \frac{n-3}{n+4} \AAG_{n-1,1}^0(z) - \frac{2n+1}{n+4} z^2 \AAG_{n,0}^0(z) .
\end{equation}
These relations determine all $\AAG_{n,0}^0(z)$ and $\AAG_{n,1}^0(z)$,
from which one can construct recursively all $\AAG_{n,n'}^m(z)$, as
described above.

In the limit of a disk ($z = \sinh\alpha_0 \to 0$), one gets
$\AAG_{0,0}^0(0) = 1$ and $\AAG_{1,1}^0(0) = 1/2$, from which all
other elements $\AAG_{n,n'}^m(0)$ can be recursively constructed.




\subsection{Orthogonality of eigenfunctions}
\label{sec:orthogonality}

Let us check that the eigenfunctions $v_k$ obtained from the
eigenvectors $\VV_k$ of the matrix $\MM$ are orthogonal to each other.
For oblate spheroids, we have
\begin{align}  \nonumber
& (v_{k_1},v_{k_2})_{L^2(\pa)} = \int\limits_{-\pi/2}^{\pi/2} d\theta \int\limits_0^{2\pi} d\phi \,
h_\theta \, h_\phi \, v_{k_1}^* \, v_{k_2} \\  \nonumber
& = a_E^2 \cosh \alpha_0 \sum\limits_{m_1,n_1,m_2,n_2} [\VV_{k_1}^*]_{m_1n_1} [\VV_{k_2}]_{m_2n_2} \\ \nonumber
& \times  \bar{G}_{m_1n_1,m_2n_2}(\sinh\alpha_0) \\
\label{eq:auxil14}
& = a_E^2 \cosh \alpha_0 \frac{1}{\mu_{k_1}} \sum\limits_{m_2,n_2}  [\VV_{k_1}^*]_{m_2n_2} c_{m_2n_2} [\VV_{k_2}]_{m_2n_2} , 
\end{align}
where we used Eqs. (\ref{eq:auxil12}, \ref{eq:auxil13}), and
$\bar{G}_{m_1n_1,m_2n_2}(\sinh\alpha_0)$ are given by
Eq. (\ref{eq:Gmn0}).  It remains to show that the above sum vanishes
when $\mu_{k_1} \ne \mu_{k_2}$.  

For this purpose, we rewrite Eq. (\ref{eq:M_oblate}) in a matrix form
as $\MM = \FF \cc$, where $\FF$ is the symmetric matrix with elements
\begin{equation}
\bar{F}_{mn,m'n'}(\sinh\alpha_0) = 2\pi\delta_{m,m'} a_{mn} a_{m'n'} \bar{F}_{n,n'}^m(\sinh\alpha_0) ,
\end{equation} 
and $\cc$ is the diagonal matrix formed by $c_{mn}$.  On one hand, one
can transpose the relation $\MM \VV_{k_1} = \mu_{k_1} \VV_{k_1}$ and
multiply it by $\cc \VV_{k_2}$ on the right to get
\begin{equation*}
\VV_{k_1}^\dagger \cc \FF \cc \VV_{k_2} = \mu_{k_1} \VV_{k_1}^\dagger \cc \VV_{k_2} .
\end{equation*}
On the other hand, multiplication of the relation $\MM \VV_{k_2} =
\mu_{k_2} \VV_{k_2}$ by $\VV_{k_1}^\dagger \cc$ on the left yields
\begin{equation*}
\VV_{k_1}^\dagger \cc \FF \cc \VV_{k_2} = \mu_{k_2} \VV_{k_1}^\dagger \cc \VV_{k_2}.
\end{equation*}
Subtracting these equations, one gets
\begin{equation*}
(\mu_{k_1} - \mu_{k_2}) \VV_{k_1}^\dagger \cc \VV_{k_2} = 0,
\end{equation*}
that implies the orthogonality for any $\mu_{k_1} \ne \mu_{k_2}$:
\begin{equation}  \label{eq:Vk_ortho}
\VV_{k_1}^\dagger \cc \VV_{k_2} = \sum\limits_{m,n} [\VV_{k_1}^*]_{mn} c_{mn} [\VV_{k_2}]_{mn} = 0,
\end{equation}
and thus the orthogonality of the eigenfunctions $v_{k_1}$ and
$v_{k_2}$ due to Eq. (\ref{eq:auxil14}).  

Since the matrix $\MM$ is not symmetric, its eigenvectors $\VV_k$ are
not orthogonal to each other; in fact, their orthogonality relation
(\ref{eq:Vk_ortho}) includes the weighting coefficient $c_{mn}$.  For
this reason, it is more convenient to consider rescaled eigenvectors
$\tilde{\VV}_k = \cc^{\frac12} \VV_k$, which are the eigenvectors of
the Hermitian matrix
\begin{equation}
\tilde{\MM} =  \cc^{\frac12} \MM \cc^{-\frac12} = \cc^{\frac12} \FF \cc^{\frac12}  .
\end{equation}
In fact, one has $\tilde{\MM} \tilde{\VV}_k = \mu_k \tilde{\VV}_k$,
with the same eigenvalue $\mu_k$.  The rescaled eigenvectors
$\tilde{\VV}_k$ are orthogonal to each.

Note that Eq. (\ref{eq:auxil14}) allows one to compute the
$L^2(\pa)$-norm of each eigenfunction directly from the corresponding
eigenvector:
\begin{align}  \nonumber
& \|v_{k}\|_{L^2(\pa)}^2 = \int\limits_{-\pi/2}^{\pi/2} d\theta \int\limits_0^{2\pi} d\phi \,
h_\theta \, h_\phi \, |v_k|^2 \\  \nonumber
& = a_E^2 \cosh \alpha_0 \frac{1}{\mu_k} \sum\limits_{m,n}  [\VV_k^*]_{mn} c_{mn} [\VV_k]_{mn} \\  \label{eq:vknorm_oblate}
& = a_E^2 \cosh \alpha_0 \frac{\VV_k^\dagger \cc \VV_k }{\mu_k} = a_E^2 \cosh \alpha_0 \frac{\tilde{\VV}_k^\dagger \tilde{\VV}_k }{\mu_k} \,.
\end{align}
This straightforward computation helps to avoid numerical integration
over $\theta$.  A similar computation for prolate spheroids yields:
\begin{equation}   \label{eq:vknorm_prolate}
\|v_{k}\|_{L^2(\pa)}^2 = a_E^2 \sinh \alpha_0 \frac{\VV_k^\dagger \cc \VV_k }{\mu_k}
= a_E^2 \sinh \alpha_0 \frac{\tilde{\VV}_k^\dagger  \tilde{\VV}_k }{\mu_k} \,.
\end{equation}

In the same vein, one can compute the projection of an eigenfunction
$v_k$ onto a constant.  For oblate spheroids, one gets
\begin{align}  \nonumber
(1,v_k)_{L^2(\pa)} & = \int\limits_{-\pi/2}^{\pi/2} d\theta \int\limits_0^{2\pi} d\phi\, h_\theta \, h_\phi \, v_k \\  \nonumber
& = a_E^2 \cosh \alpha_0  \sum\limits_{m,n} [\VV_k]_{mn} \bar{G}_{mn,00}(\sinh\alpha_0) \sqrt{4\pi} \\  \label{eq:vk_1_oblate}
& = \sqrt{4\pi} \, a_E^2 \cosh \alpha_0 \frac{c_{00}}{\mu_k} [\VV_k]_{00} ,
\end{align}
where we used that $\bar{Y}_{00} = a_{00} = 1/\sqrt{4\pi}$.  For
instance, for the exterior problem, substitution of $c_{00}$ from
Eq. (\ref{eq:cmn_oblate}) yields
\begin{equation}
(1,v_k)_{L^2(\pa)} =  \frac{\sqrt{4\pi} a_E}{iQ_0(i\sinh\alpha_0)} \, \frac{[\VV_k]_{00}}{\mu_k} \,.
\end{equation}

For prolate spheroids, a similar computation gives
\begin{equation}   \label{eq:vk_1_prolate}
(1,v_k)_{L^2(\pa)} = \sqrt{4\pi} \, a_E^2 \sinh \alpha_0  \frac{c_{00}}{\mu_k} [\VV_k]_{00} .
\end{equation}
For instance, for the exterior problem, substitution of $c_{00}$ from
Eq. (\ref{eq:cmn_prolate}) yields
\begin{equation}   \label{eq:vk_1_prolateE}
(1,v_k)_{L^2(\pa)} =  \frac{\sqrt{4\pi} \, a_E}{Q_0(\cosh \alpha_0)}\, \frac{[\VV_k]_{00}}{\mu_k}  .
\end{equation}

\subsection{Axisymmetric setting}
\label{sec:axi}

In many applications, it is sufficient to look at axisymmetric
eigenfunctions, which do not depend on the angle $\phi$ and thus
correspond to $m = 0$.  These eigenfunctions can be constructed by
diagonalizing the matrix $\MM_0$, for which the computation can be
further simplified.  We briefly discuss this case below for prolate
and oblate spheroids.

\subsubsection*{Prolate spheroids}
\label{sec:axiP}

In this case, Eq. (\ref{eq:f_Ymn}) is reduced to
\begin{equation}
f(\theta) = \sum\limits_{n=0}^\infty f_n \, \psi_n(\theta),
\end{equation}
where
\begin{equation}
\psi_n(\theta) = \sqrt{n+1/2} \, P_n(\cos\theta)
\end{equation}
are normalized Legendre polynomials.  The action of $\M$ reads then
\begin{equation}
[\M f](\theta) = \sum\limits_{n=0}^\infty [\MM_0 f]_n \psi_n(\theta) ,
\end{equation}
where the (infinite-dimensional) matrix $\MM_0$ represents the operator
$\M$ on the orthonormal basis of Legendre polynomials, with
\begin{align}   \label{eq:M_prolateA} 
[\MM_0]_{n,n'} & = \MM_{0n,0n'}  \\  \nonumber
& = \sqrt{(n+1/2)(n'+1/2)} \, c_{0n'} \, F_{n,n'}^0(\cosh\alpha_0) ,
\end{align}
where
\begin{equation}  \label{eq:bn_prolate}
c_{0n} = - \frac{\sinh \alpha_0 \, Q'_n(\cosh \alpha_0)}{a_E Q_n(\cosh\alpha_0)} \,,
\end{equation}
and
\begin{equation}
F_{n,n'}^0(z) = \int\limits_{-1}^1 dx \frac{P_n(x) P_{n'}(x)}{\sqrt{z^2 - x^2} }  \,.
\end{equation}
By symmetry, the elements $F_{n,n'}^0(z)$ are zero when $n+n'$ is odd.
Once the eigenvectors $\VV_{0n}$ of the matrix $\MM_0$ are found, the
eigenfunction $v_{0n}$ and the Steklov eigenfunction $V_{0n}$ are
given by
\begin{equation}  \label{eq:vk_prolateA}
v_{0n}(\theta) = \sum\limits_{n'=0}^\infty [\VV_{0n}]_{0n'} \psi_{n'}(\theta)
\end{equation} 
and
\begin{equation}  \label{eq:Vk_prolateA}
V_{0n}(\alpha,\theta) = \sum\limits_{n'=0}^\infty 
\frac{Q_{n'}(\cosh \alpha)}{Q_{n'}(\cosh \alpha_0)} [\VV_{0n}]_{0n'}  \psi_{n'}(\theta).
\end{equation}

\subsubsection*{Oblate spheroids}

In the same vein, one employs the normalized Legendre polynomials
\begin{equation}
\psi_n(\theta) = \sqrt{n+1/2}\, P_n(\sin\theta)
\end{equation}
to get the matrix representation of $\M$:
\begin{equation}  \label{eq:M_oblateA}
[\MM_0]_{0n,0n'} = \sqrt{(n+1/2)(n'+1/2)} \, c_{0n'}\, i F_{n,n'}^0(i\sinh\alpha_0) ,
\end{equation}
where
\begin{equation}  \label{eq:bn_oblate}
c_{0n} = \frac{\cosh \alpha_0 Q'_n(i\sinh \alpha_0)}{i a_E Q_n(i\sinh\alpha_0)} \,.
\end{equation}

\vskip 3mm
\subsubsection*{Disk}

In the limit $a = 0$, the coefficients $c_{0n}$ from
Eq. (\ref{eq:bn_oblate}) approach constants; in fact, using
\begin{align}
Q_{2n}(0) & = (-1)^n \frac{(2n)!}{4^n \, (n!)^2}  \, \frac{\pi}{2i} \,,  \\  
Q_{2n-1}(0) & = (-1)^n \frac{2^{2n-1} (n!)^2}{n (2n)!} \,,
\end{align}
and the recurrence relation (\ref{eq:dQn}) to evaluate $Q'_n(0) = n
Q_{n-1}(0)$, we deduce Eq. (\ref{eq:c0n_disk}).  Note that these
coefficients can be found iteratively as
\begin{equation}
c_{0n} = \frac{c_{0(n-2)}}{(1 - 1/n)^2} \,.
\end{equation}
One has $c_{0n} \simeq n$ as $n$ increases.


\begin{thebibliography}{10}


\bibitem{Cheney99}			M. Cheney, D. Isaacson, and J. C. Newell,
					Electrical Impedance Tomography,
					SIAM Rev. {\bf 41}, 85-101 (1999).

\bibitem{Calderon80}			A. P. Calder\'on, 
					On an inverse boundary value problem, 
					Seminar on Numerical Analysis and its Applications to Continuum Physics, 
					Soc. Brasileira de Matem\'atica, R\'io de Janeiro, (1980), 65-73;
					Reprinted in Comput. Appl. Math. {\bf 25}, 2-3, 133-138 (2006).

\bibitem{Borcea02}			L. Borcea,
					Electrical impedance tomography,
					Inv. Prob. {\bf 18}, R99 (2002).

\bibitem{Sylvester87}			J. Sylvester and G. Uhlmann,
					A Global Uniqueness Theorem for an Inverse Boundary Value Problem,
					Ann. Math. {\bf 125}, 153-169 (1987).

\bibitem{Curtis91}			E. Curtis and J. Morrow,
					The Dirichlet to Neumann map for a resistor network,
					SIAM J. Appl. Math. {\bf 51}, 1011-1029 (1991).


\bibitem{Agranovich}			M. S. Agranovich, B. Z. Katsenelenbaum, A. N. Sivov, and N. N. Voitovich,
					{\it Generalized Method of Eigenoscillations in Diffraction Theory}
					(Wiley-VCH, Berlin, 1999).

\bibitem{Smith96}			B. F. Smith, 
					Domain Decomposition Methods for Partial Differential Equations,
					in ``Parallel Numerical Algorithms'', Eds. D. E. Keyes, A. Sameh, V. Venkatakrishnan 
					(Springer, 1996), pp. 225-243. 

\bibitem{Givoli98}			D. Givoli, I. Patlashenko, and J. B. Keller,
					Discrete Dirichlet-to-Neumann maps for unbounded domains,
					Comput. Methods Appl. Mech. Engrg. {\bf 164}, 173-185 (1998).

\bibitem{Delitsyn18}			A. Delitsyn and D. S. Grebenkov, 
					Mode matching methods in spectral and scattering problems, 
					Quart. J. Mech. Appl. Math. {\bf 71}, 537-580 (2018).

\bibitem{Delitsyn22}			A. Delitsyn and D. S. Grebenkov, 
					Resonance scattering in a waveguide with identical thick perforated barriers, 
					Appl. Math. Comput. {\bf 412}, 126592 (2022).






\bibitem{Auchmuty04}			G. Auchmuty,
					Steklov eigenproblems and the representation of solutions of 
					elliptic boundary value problems,
					Numer. Funct. Anal. Optim. {\bf 25}, 321-348 (2004).

\bibitem{Auchmuty13}			G. Auchmuty and Q. Han,
					Spectral representations of solutions of linear elliptic equations on exterior regions,
					J. Math. Anal. Appl. {\bf 398}, 1-10 (2013).

\bibitem{Auchmuty14}			G. Auchmuty and Q. Han,
					Representations of Solutions of Laplacian Boundary Value Problems on Exterior Regions,
					Appl. Math. Optim. {\bf 69}, 21-45 (2014).  

\bibitem{Auchmuty15}			G. Auchmuty and M. Cho, 
					Boundary integrals and approximations of harmonic functions, 
					Numer. Funct. Anal. Optim. {\bf 36}, 687-703 (2015).

\bibitem{Auchmuty18}			G. Auchmuty,
					Steklov Representations of Green's Functions for Laplacian Boundary Value Problems,
					Appl. Math. Optim. {\bf 77} (2018).





\bibitem{Grebenkov06}			D. S. Grebenkov, M. Filoche, and B. Sapoval, 
					Mathematical Basis for a General Theory of Laplacian Transport towards Irregular Interfaces, 
					Phys. Rev. E {\bf 73}, 021103 (2006). 

\bibitem{Grebenkov19}			D. S. Grebenkov,
					Spectral theory of imperfect diffusion-controlled reactions on heterogeneous catalytic surfaces,
					J. Chem. Phys. {\bf 151}, 104108 (2019).

\bibitem{Grebenkov20}			D. S. Grebenkov,
					Paradigm shift in diffusion-mediated surface phenomena,
					Phys. Rev. Lett. {\bf 125}, 078102 (2020).


\bibitem{Grebenkov22a}			D. S. Grebenkov, 
					An encounter-based approach for restricted diffusion with a gradient drift, 
					J. Phys. A: Math. Theor. {\bf 55}, 045203 (2022). 





\bibitem{Grebenkov19c}			D. S. Grebenkov,
					Probability distribution of the boundary local time of reflected Brownian motion in Euclidean domains,
					Phys. Rev. E {\bf 100}, 062110 (2019).

\bibitem{Grebenkov21a}			D. S. Grebenkov, 
					Statistics of boundary encounters by a particle diffusing outside a compact planar domain, 
					J. Phys. A.: Math. Theor. {\bf 54}, 015003 (2021).

\bibitem{Grebenkov22d}			D. S. Grebenkov, 
					Statistics of diffusive encounters with a small target: Three complementary approaches, 
					J. Stat. Mech. 083205 (2022). 


\bibitem{Grebenkov20c}			D. S. Grebenkov,
					Joint distribution of multiple boundary local times and related first-passage time problems,
					J. Stat. Mech. 103205 (2020).



\bibitem{Levitin}		M. Levitin, D. Mangoubi, and I. Polterovich,
				{\it Topics in Spectral Geometry}
				(Graduate Studies in Mathematics, vol. 237; American Mathematical Society, 2023).

\bibitem{Girouard17}		A. Girouard and I. Polterovich,
				Spectral geometry of the Steklov problem, 
				J. Spectr. Th. {\bf 7}, 321-359 (2017).

\bibitem{Colbois23}		B. Colbois, A. Girouard, C. Gordon, and D. Sher,
				Some recent developments on the Steklov eigenvalue problem,
				Rev. Mat. Complut. (2023).



\bibitem{Grebenkov20b}			D. S. Grebenkov, 
					Surface Hopping Propagator: An Alternative Approach to Diffusion-Influenced Reactions, 
					Phys. Rev. E {\bf 102}, 032125 (2020).





\bibitem{Arendt15}		W. Arendt and A. F. M. ter Elst, 
				The Dirichlet-to-Neumann Operator on Exterior Domains, 
				Potential Anal. {\bf 43}, 313-340 (2015).


\bibitem{Christiansen23}	T. J. Christiansen and K. Datchev, 
				Low energy scattering asymptotics for planar obstacles, 
				Pure Appl. Anal. {\bf 5}, 767-794 (2023).

\bibitem{Xiong23}		C. Xiong, 
				Sharp bounds for the first two eigenvalues of an exterior Steklov eigenvalue problem, 
				ArXiv:2304.11297v1 (2023).





\bibitem{Morse}			P. M. Morse and H. Feshbach,
				{\it Methods of Theoretical Physics}
				(New York: McGraw-Hill, 1953).

\bibitem{Smythe}		W. R. Smythe,
				{\it Static and Dynamic Electricity}, 3rd ed.
				(New York: McGraw-Hill, 1968).


\bibitem{Barucq10}		H. Barucq, R. Djellouli, and A. Saint-Guirons,
				Three-dimensional approximate local DtN boundary conditions for prolate spheroid boundaries,
				J. Comput. Appl. Math. {\bf 234}, 1810-1816 (2010).

\bibitem{LeGia11}		Q. T. Le Gia, E. P. Stephan, and T. Tran,
				Solution to the Neumann problem exterior to a prolate spheroid by radial basis functions,
				Adv. Comput. Math. {\bf 34}, 83-103 (2011).

\bibitem{Costea11}		A Costea, Q. T. Le Gia, E. P. Stephan, and T. Tran,
				Meshless BEM and overlapping Schwarz preconditioners for exterior problems on spheroids,
				Stud. Geophys. Geod. {\bf 55}, 465-477 (2011).

\bibitem{Luo15}			X. Luo, Q. Du, and L. Liu,
				A D-N alternating algorithm for exterior 3D Poisson problem with prolate spheroid boundary,
				Appl. Math. Comput. {\bf 269}, 252-264 (2015).

\bibitem{Gomez15}		D. Gomez and A. V. Cheviakov,
				Asymptotic analysis of narrow escape problems in nonspherical three-dimensional domains,
				Phys. Rev. E {\bf 91}, 012137 (2015).

\bibitem{Xue17}			C. Xue ans S. Deng,
				Green's function and image system for the Laplace operator in the prolate spheroidal geometry,
				AIP Advances {\bf 7}, 015024 (2017).

\bibitem{Traytak18}		S. D. Traytak and D. S. Grebenkov, 
				Diffusion-influenced reaction rates for active ``sphere-prolate spheroid'' pairs and Janus dimers, 
				J. Chem. Phys. {\bf 148}, 024107 (2018).

\bibitem{Piazza19}		F. Piazza and D. S. Grebenkov, 
				Diffusion-controlled reaction rate on non-spherical partially absorbing axisymmetric surfaces, 
				Phys. Chem. Chem. Phys. {\bf 21}, 25896-25906 (2019). 

\bibitem{Chaigneau22}		A. Chaigneau and D. S. Grebenkov, 
				First-passage times to anisotropic partially reactive targets, 
				Phys. Rev. E {\bf 105}, 054146 (2022).

\bibitem{Chaigneau23}		A. Chaigneau and D. S. Grebenkov, 
				Effects of target anisotropy on harmonic measure and mean first-passage time, 
				J. Phys. A: Math. Theor. {\bf 56}, 235202 (2023).


\bibitem{Korn}			G. A. Korn and T. M. Korn,
				{\it Mathematical Handbook for Scientists and Engineers}
				(New York: McGraw-Hill, 1961).






\bibitem{Sneddon}		I. N. Sneddon, 
				{\it Mixed Boundary Value Problems in Potential Theory}
				(Wiley, NY, 1966).


\bibitem{Garnett}		J. B. Garnett and D. E. Marshall,
				{\it Harmonic Measure}
				(Cambridge University Press, 2005).



\bibitem{VanKampen}		N. G. Van Kampen,
				{\it Stochastic Processes in Physics and Chemistry}
				(Elsevier, Amsterdam, 1992).

\bibitem{Redner} 		S. Redner, 
				{\it A Guide to First Passage Processes}
				(Cambridge, Cambridge University press, 2001).





\bibitem{Schuss}		Z. Schuss, 
				{\it Brownian Dynamics at Boundaries and Interfaces in Physics, Chemistry and Biology}
				(Springer: New York, USA, 2013).

\bibitem{Metzler} 		R. Metzler, G. Oshanin, and S. Redner (Eds), 
				{\it First-Passage Phenomena and Their Applications}
				(Singapore, World Scientific, 2014).

\bibitem{Lindenberg}		K. Lindenberg, R. Metzler, and G. Oshanin (Eds),
				{\it Chemical Kinetics: Beyond the Textbook}
				(World Scientific, New Jersey, 2019).


\bibitem{North66}		A. M. North,  
				Diffusion-controlled reactions,
				Q. Rev. Chem. Soc. {\bf 20}, 421--440 (1966).

\bibitem{Wilemski73}		G. Wilemski and M. Fixman,
				General theory of diffusion-controlled reactions, 
				J. Chem. Phys. {\bf 58}, 4009--4019 (1973).

\bibitem{Calef83}		D. F. Calef and J. M. Deutch, 
				Diffusion-Controlled Reactions,
				Ann. Rev. Phys. Chem. {\bf 34}, 493--524 (1983).

\bibitem{Berg85}		O. G. Berg and P. H. von Hippel,  
				Diffusion-Controlled Macromolecular Interactions,
				Ann. Rev. Biophys. Biophys. Chem. {\bf 14}, 131--160 (1985).

\bibitem{Rice85}		S. A. Rice, 
				{\it Diffusion-limited reactions} 
				(Elsevier, Amsterdam, 1985).

\bibitem{Bressloff13}		P. C. Bressloff and J. Newby,
				Stochastic models of intracellular transport,
				Rev. Mod. Phys. {\bf 85}, 135-196 (2013).

\bibitem{Benichou14}		O. B\'enichou and R. Voituriez,
				From first-passage times of random walks in confinement to geometry-controlled kinetics,
				Phys. Rep. {\bf 539}, 225-284 (2014).

\bibitem{Grebenkov23n}		D. S. Grebenkov,
				Diffusion-Controlled Reactions: An Overview,
				Molecules {\bf 28}, 7570 (2023).



\bibitem{Smoluchowski17}	M. Smoluchowski, 
				Versuch einer Mathematischen Theorie der Koagulations Kinetic Kolloider L\"osungen, 
				Z. Phys. Chem. {\bf 92U}, 129-168 (1918).

\bibitem{Collins49}			F. C. Collins and G. E. Kimball, 
					Diffusion-controlled reaction rates, 
					J. Colloid Sci. {\bf 4}, 425-437 (1949).


\bibitem{Grebenkov18b}          D. S. Grebenkov and D. Krapf, 
				Steady-state reaction rate of diffusion-controlled reactions in sheets, 
				J. Chem. Phys 149, 064117 (2018).




\bibitem{Grebenkov23}		D. S. Grebenkov, 
				Encounter-based approach to the escape problem, 
				Phys. Rev. E {\bf 107}, 044105 (2023). 


\bibitem{Andreev04}		A. B. Andreev and T. D. Todorov, 
				Isoparametric finite-element approximation of a Steklov eigenvalue problem, 
				IMA J. Num. Anal. {\bf 24}, 309-322 (2004).

\bibitem{Yang09}		Y. Yang, Q. Li, and S. Li, 
				Nonconforming finite element approximations of the Steklov eigenvalue problem, 
				Appl. Num. Math. {\bf 59}, 2388-2401 (2009).

\bibitem{Bi11}			H. Bi, Y. Yang, 
				A two-grid method of the non-conforming Crouzeix-Raviart element for the Steklov eigenvalue problem,
				Appl. Math. Comput. {\bf 217}, 9669-9678 (2011).

\bibitem{Li11}			Q. Li and Y. Yang, 
				A two-grid discretization scheme for the Steklov eigenvalue problem, 
				J. Appl. Math. Comput. {\bf 36}, 129-139 (2011).

\bibitem{Li13}			Q. Li, Q. Lin, and H. Xie, 
				Nonconforming finite element approximations of the Steklov eigenvalue problem and its lower bound approximations, 
				Appl. Math. {\bf 58}, 129-151 (2013).

\bibitem{Xie14}			H. Xie, 
				A type of multilevel method for the Steklov eigenvalue problem, 
				IMA J. Num. Anal. {\bf 34}, 592-608 (2014).

\bibitem{Bi16}			H. Bi, H. Li, and Y. Yang, 
				An adaptive algorithm based on the shifted inverse iteration for the Steklov eigenvalue problem,
				Appl. Num. Math. {\bf 105} 64-81 (2016).

\bibitem{Bogosel16}		B. Bogosel, 
				The method of fundamental solutions applied to boundary eigenvalue problems, 
				J. Comput. Appl. Math. {\bf 306}, 265-285 (2016).

\bibitem{Akhmetgaliyev17}	E. Akhmetgaliyev, C.-Y. Kao, and B. Osting, 
				Computational methods for extremal Steklov problems, 
				SIAM J. Contr. Optim. {\bf 55}, 1226-1240 (2017).

\bibitem{Alhejaili19}		W. Alhejaili and C.-Y. Kao, 
				Numerical studies of the Steklov eigenvalue problem via conformal mappings, 
				Appl. Math. Comput. {\bf 347}, 785-802 (2019).

\bibitem{Chen20}		J.-T. Chen, J.-W. Lee, and K.-T. Lien, 
				Analytical and numerical studies for solving Steklov eigenproblems by using the
				boundary integral equation method/boundary element method, 
				Engnr. Anal. Bound. Elem. {\bf 114}, 136-147 (2020).

\bibitem{Chaigneau24}		A. Chaigneau and D. S. Grebenkov,
				A numerical study of the Dirichlet-to-Neumann operator in planar domains 
				(submitted). 




\bibitem{Richter74}		P. H. Richter and M. Eigen,
				Diffusion controlled reaction rates in spheroidal geometry: 
				Application to repressor-operator association and membrane bound enzymes,
				Biophys. Chem. {\bf 2}, 255-263 (1974).

\bibitem{Grimes18}		D. R. Grimes and F. J. Currell,
				Oxygen diffusion in ellipsoidal tumour spheroids,
				J. R. Soc. Interface {\bf 15}, 20180256 (2018).

\bibitem{Gadzinowski21}		M. Gadzinowski, D. Mickiewicz, and T. Basinska,
				Spherical versus prolate spheroidal particles in biosciences: Does the shape make a difference?
				Polym. Adv. Technol. 1-10 (2021).




\bibitem{Bressloff22a}			P. C. Bressloff,
					Diffusion-mediated absorption by partially-reactive targets: 
					Brownian functionals and generalized propagators,
					J. Phys. A: Math. Theor. {\bf 55}, 205001 (2022).

\bibitem{Bressloff22b}			P. C. Bressloff,
					Narrow capture problem: an encounter-based approach to partially reactive targets,
					Phys. Rev. E {\bf 105}, 034141 (2022).

\bibitem{Bressloff22c}			P. C. Bressloff, 
					A probabilistic model of diffusion through a semipermeable barrier,
					Proc. Roy. Soc. A {\bf 478}, 20220615 (2022).

\bibitem{Bressloff22d}			P. C. Bressloff,
					Diffusion-mediated surface reactions and stochastic resetting,
					J. Phys. A: Math. Theor. {\bf 55}, 275002 (2022).



\bibitem{Abramowitz}  		M. Abramowitz and I. A. Stegun,
				{\it Handbook of Mathematical Functions with Formulas, Graphs, and Mathematical Tables}
				(United States Department of Commerce, National Bureau of Standards, 1964).




\end{thebibliography}
\end{document}